%% file: main.tex
\documentclass{aastex62}

\newcommand{\angstrom}{\mbox{\normalfont\AA}}

\newcommand{\esbupdate}[1]{{\normalfont #1}}
\newcommand{\cmmupdate}[1]{{\normalfont #1}}

\newcommand{\CCSN}{CCSN}
\newcommand{\CCSNe}{CCSNe}

\newcommand{\Mbounce}{M_{bounce}}
\newcommand{\Mprog}{M_{prog}}
\newcommand{\Msol}{M_{\odot}}
\newcommand{\Mrem}{M_{rem}}
\newcommand{\Eexp}{E_{exp}}
\newcommand{\Minj}{M_{inj}}
\newcommand{\Einj}{E_{inj}}
\newcommand{\tinj}{t_{inj}}
\newcommand{\cbar}{\Tilde{c}}
\newcommand{\fancylet}[1]{{\mathcal{#1}}}

\newcommand{\carb}{\mathrm{C}}
\newcommand{\sili}{\mathrm{Si}}
\newcommand{\SiC}{\mathrm{SiC}}
\newcommand{\TiC}{\mathrm{TiC}}
\newcommand{\silica}{\mathrm{SiO}_{2}}
\newcommand{\magnesia}{\mathrm{MgO}}
\newcommand{\forsterite}{\mathrm{Mg}_{2}\mathrm{SiO}_{4}}
\newcommand{\enstatite}{\mathrm{MgSiO}_{3}}
\newcommand{\alumina}{\mathrm{Al}_{2}\mathrm{O}_{3}}
\newcommand{\magnetite}{\mathrm{Fe}_{3}\mathrm{O}_{4}}

\newcommand{\uT}{\mathrm{K}}

\newcommand{\uerg}{\mathrm{erg}}

\newcommand{\es}[2]{#1\times10^{#2}}
\newcommand{\tens}[1]{10^{#1}}

\usepackage{float}
\usepackage{amsmath}
\usepackage{amssymb}
\usepackage{todonotes}
\usepackage[normalem]{ulem}

\usepackage{lineno}

\graphicspath{{./}{Images/}}

\received{March 12, 2021}
\revised{July 19, 2021}

\submitjournal{ApJ}

\shorttitle{Dust in SN Explosions}
\shortauthors{Brooker et al.} 

\begin{document}

\title{Dependence of Dust Formation on the Supernova Explosion}

\author[0000-0001-7404-4100]{Ezra S. Brooker}
\affil{Florida State University, Tallahassee, FL, 32306, USA}
\collaboration{(Los Alamos National Laboratory)}

\author[0000-0001-5570-6666]{Sarah M. Stangl}
\affiliation{University of Oklahoma}
\collaboration{(Los Alamos National Laboratory)}

\author[0000-0002-7827-2247]{Christopher M. Mauney}
\affiliation{Center for Theoretical Astrophysics, Los Alamos National Laboratory,
  Los Alamos, NM, 87545, USA}
  
\author[0000-0003-2624-0056]{C.~L. Fryer}
\affiliation{Center for Theoretical Astrophysics, Los Alamos National Laboratory,
  Los Alamos, NM, 87545, USA}
\affiliation{Computer, Computational, and Statistical Sciences Division,
  Los Alamos National Laboratory, Los Alamos, NM, 87545, USA}
\affiliation{The University of Arizona, Tucson, AZ 85721, USA}
\affiliation{Department of Physics and Astronomy, The University of New Mexico, Albuquerque, NM 87131, USA}
\affiliation{The George Washington University, Washington, DC 20052, USA}

\begin{abstract}

We investigate the properties, composition, and dynamics of dust formation and growth for a diverse set of core-collapse supernovae (\CCSNe), varying the progenitor mass, explosion energy, and engine type. These explosions are evolved with a 1-D Lagrangian hydrodynamics code out to several hundred days to model the ejecta as it expands and cools. A multigrain dust nucleation and growth model is applied to these results. We find that higher explosion energies lead to an earlier onset of dust formation, smaller grain sizes, and larger silicate abundances. Further, we see that nuclear burning during the explosion leads to enhanced formation of silicate dust. Finally, we build composite models from our suite to predict the efficiency of \CCSNe\ dust production as a function of metallicity.
\end{abstract}

\keywords{Supernova, core-collapse supernova, dust formation, Lagrangian hydrocode}

\section{Introduction} \label{sec:intro}

Dust is ubiquitous in the interstellar medium (ISM) of most galaxies, typically in the form of carbonaceous or silicate cores with a mantle of accumulated ices. \cmmupdate{A vital component of stellar and  galactic life-cycles, understanding the formation of dust in different galactic environments is key to understanding the evolution of those host galaxies. Dust cools and insulates collapsing molecular clouds, allowing for more efficient star formation. Heavy elements are locked into dust grains, depleting the gas phase of these elements. In the ISM the dust grains provide a site for the formation of H$_2$O and other complex molecules through diffusion on the surface. \citep{draine2003interstellar}. 
}

\cmmupdate{Of the many theoretical explorations of astrophysics, the transformation of stellar vapor to interstellar molecular fog is shrouded in conjecture and guess-work. While there is a substantial body of work that covers terrestrial phenomenon, the formation, growth, and evolution of large molecular particles in the near-vacuum of the environment remains a subject of intense conjecture in the stellar and galactic background. We seek to understand the dynamics of time and distance scales over several magnitudes ranging from the chemical reactions at the molecular scale to dynamical reactions at the stellar scale, and then back again. The scope of this study is difficult to simply encompass.
}
\cmmupdate{There have been several attempts to this extremely difficult question.}
\cmmupdate{Dust formation and processing has been observed in stellar winds, \CCSNe, and the atmospheres of AGB stars. Dust from Type Ia SNe has also be suggested as a significant contributor \citep{gomez2012dust}.
There are also several investigations into the cold formation and growth of dust in molecular clouds \citep{marin2020low,mattsson2020grain}. The prime producer of galactic dust must have changed over time, as galaxies at high-$z$ lack late-age producers of dust in the local galactic environment such as AGB stars and Type Ia SNe. If dust formed in the outflows of \CCSNe\ is a significant source of ISM dust, then production in low-metallicity galaxies of the early universe is likely to be dominated by \CCSNe\ \citep{sadavoy2019life}.
}

\cmmupdate{Observations of local \CCSNe\, such as 1987A \citep{Dwek_2015,Wesson2014,wesson2015timing} and Cassiopeia A \citep{Arendt_2014,Priestley2019MNRAS.485..440P}, show abundant dust masses in \CCSNe\ ejecta prior to interacting with the ISM. Heating from post-explosion shocks will disrupt grain formation and growth, and it is argued that this will prevent any significant portion of dust grains formed in the outflow from surviving long enough to reach into the ISM \citep{bocchio2016dust}. However, dust material has been seen to survive and reform after the passing of forward shocks in the ejecta of 1987A \citep{matsuura2019sofia}. Large dust grains capable of surviving shock destruction have been seen in abundance in SN2010jl \citep{gall2014rapid}
Models of dust formation across cosmic time complicate this picture further, with an epoch of formation in stellar ejecta at $z  < 2 $, later overtaken by growth in the ISM as the main channel of dust production up to the present \citep{triani2020origin}.}

The explosion energy, explosive engine, metallicity, and progenitor mass of the \CCSN\ will all impact the subsequent dust formation history and composition \citep{muller2016simple}.  Observations of the supernova ejecta probe the detailed composition of the ejecta which, in turn, can be used to probe the properties of the progenitor star and the process of the explosion. After shock breakout, the outflow expands and cools with ionized plasma recombining into the gas phase. Gas-phase reactions occur and change the initial abundance of free gas phase species into a rich mixture of compounds \citep{10.1093/mnras/sty2060}. At densities and temperatures starting at roughly 5000\,K, condensation nuclei (refractory dust grains) form from the free species available in the mixture. These dust grains can be spectrally observed, serving as a probe for nucleosynthetic yields and morphological tracers related to the explosion inside of the star.

Dust yields from progenitors with different masses and metallicities are still under active study. Understanding how \CCSNe\ produce different dust properties and compositions can inform astrophysicists about stellar and galactic evolution. Dust formation studies of a limited number of \CCSN\ progenitor configurations have previously been undertaken covering various contexts.  For example, studies have been done of Population III stars \citep{Nozawa_2013} and the effects of metallicity and stellar rotation \citep{marassi2019supernova}. \cmmupdate{Molecule and dust precursor evolution across stellar masses was investigated in \citep{sarangi2013chemically}. This work also used Ni$_{56}$ as a proxy for looking through explosion energies.} 

In this work, we extend these previous results to include \cmmupdate{profiles and yields from high-fidelity CCSNe simulations as our starting point for hydrodynamic and dust formation evolution. This suite of CCNSe includes yields of elements formed during the collapse and bounce phase of the explosion, offering more precise initial conditions. We also use active hydrodynamics to generate a more realistic temperature and density background, as well as to incorporate the thermal and compression effects of shocks arising from the explosion.}

This paper is constructed as follows. In section \ref{sec:methods}, we describe how our \CCSNe\ models are constructed, as well as the dust model we use. Section \ref{sec:results} describes our results, including the distribution of dust in the ejecta, the composition, mass, and size of dust grains, and a comparison of different progenitor types. Finally, in section \ref{sec:conclusions} we discuss our conclusions and suggest observational applications.



\section{Methods} \label{sec:methods}

\cmmupdate{Chemical activity in the ejecta environment is controlled by composition, temperature, and pressure. Our entry point into modeling this environment is an initial 1-D profile of a \CCSNe immediately prior to shock breakout.
We proceed to map these profiles onto an extended 1-D grid extending out to a presumed terminus at the ISM, and append a stellar wind from the stars surface to the boundary.
The exploding star plus wind system is then hydrodynamically evolved out to several years, enough time that all nucleation activity will cease.
Using the density and temperature trajectories of these simulations, we then calculate the dust nucleation and growth histories for each grid cell.}

\subsection{Core-Collapse Supernova Models}

To model dust in CCSN ejecta we utilized a suite of 1D CCSN explosion models from \cite{fryer2018,andrews2020} covering a range of explosion energies ($10^{51} - 10^{53}\ \uerg$), progenitor star masses (15, 20, and 25 $\Msol$), and nucleosynthetic isotopic yields.  These calculations used a 1-dimensional core-collapse supernova code~\citep{herant94,fryer99}, \esbupdate{ referred to as FR99 from here}, which includes a gray flux-limited diffusion scheme following 3-neutrino flavors (electron, anti-electron, and $\mu$ plus $\tau$ neutrinos), a blend of equations of state to cover nuclear densities down to an ideal gas equation of state for low densities.  Nuclear burning is included using a nuclear statistical equilibrium treatment at high temperatures and a small 17-isotope network at lower temperatures.  Explosions are driven by injecting additional energy mixed into a pre-determined convective region.  

The total energy and nature of this injection (sudden energy source as expected from the convective engine versus a prolonged source produced by a magnetar or fallback accretion) are varied to produce a broad range of explosion properties. The suite of progenitor masses and explosion energies are listed in \esbupdate{ Table~\ref{tab:fryer-models} with model name designations given.} The velocity and composition of the ejecta depends both on the progenitor mass and its explosion energy.  Figure \ref{fig:abund} shows an example of the initial compositions of two 15, 20, and 25 $M_{\odot}$ progenitor models. 

\input{Figures/abundances.tex}

\subsection{Late-Time Evolution}

To follow the explosion to late times, we remove the compact core from our simulation and place the outflow onto a mesh extending out to $2.5 \times 10^{19}$ cm. Starting from the surface of the star we add a wind profile.  The winds for our different progenitors use the formulation from \citet{villata1992radiation} for a wind profile

\begin{equation}\label{eq:wind_profile}
    \dot{M}_{\rm wind}=1.2 \left( \frac{D^{\delta}\dot{M}^{CAK}_{\alpha}}{1+\alpha}\right)^{1 /(\alpha-\delta)}
\end{equation}
where $D$ and $\dot{M}_{CAK}$ are

\begin{equation}\label{eq:wind_profile_D}
\begin{split}
    D = \left(\frac{1 + Z_{He}Y_{He}}{1 + 4Y_{He}}\right)\left(\frac{9.5 \times 10^{-11}}{\pi m_{H}R_{*}^2 v_{\infty}}\right)\\
    \dot{M}_{CAK} = \frac{4 \pi G M_{*} \alpha}{\sigma_{E}v_{th}}\left[ k \Gamma \left(\frac{1-\alpha}{1 - \Gamma}\right)^{1-\alpha}\right]
\end{split}
\end{equation}
where $Z_{He}$ is the free electrons from helium, $Y_{He}$ is the helium number abundance w.r.t. $H$, $m_{H}$ is the mass of hydrogen ion, $\sigma_{E}$ is Thomspon scattering absorption coefficient per mass density, $\Gamma = L/L_{E}$ is the ratio of stellar to Eddington luminosity, $v_{\infty}$ is the escape velocity, $v_{th}$ is thermal velocity, and $k$ is a force multiplier. With a $\beta$ velocity law

\begin{equation} \label{eq:wind_profile_fin}
\begin{split}
    v(r) = v_{\infty}\left(1 - \frac{r_0}{r}\right)^{\beta}\\
    \beta = 0.95 \alpha + \frac{0.008}{\delta} + \frac{0.032v_{esc}}{500}
\end{split}
\end{equation}
where $v_{esc}$ is the escape velocity, in km s$^{-1}$.  For the wind parameters $k$, $\alpha$, $\delta$, we use the typical values:  
0.17, 0.59, 0.09 respectively.  



The corresponding density profile of the wind must include a transition from the stellar surface to the canonical $r^{-2}$ profile expected in constant velocity winds.  With our wind velocity ($v(r)$) and mass loss rate ($\dot{M}_{CAK}$), we can calculate the wind density assuming mass conservation:
\begin{equation} \label{eq_windprofile}
    \rho_{\rm wind}(r)=\dot{M}_{CAK}/(4 \pi r^2 v(r))
\end{equation}
We determine the specific energy by assuming a constant entropy wind profile. When the wind density drops below the interstellar medium density (we use a canonical value of $2.09\times 10^{-24}\ g/cm^{3}$), the density is set to the interstellar medium density. 

The subsequent late-time evolution is calculated by mapping the explosion from our core-collapse calculations into this wind density profile using a grid of 2048 Lagrangian zones.  We then follow the explosion using a simplified version of our core-collapse supernova code (FR99, without neutrino transport or equations of state for dense matter).  

Figure \ref{fig:profTrhoVel} shows the evolution of the velocity, temperature and density profiles for a model with progenitor mass of 15$\Msol$ and explosion energy $\Eexp$ of 1.69 FoE (designated M15bE1.69 in Table \ref{tab:fryer-models}) at a range of times after the launch of the explosion.  These calculations are typically evolved out to 1157 days.  The jump in the density and temperature coincides with the shock front and is reasonably well fit by the strong shock solution~\citep{1959flme.book.....L}.  As the shock propagates through the wind medium, we can see both the deceleration (comparing the velocity profiles at different times), and the subsequent reverse shock formed by this deceleration. Although we do not consider the destruction of grains from this reverse shock in this paper, our calculations provide the data to do so and this will be studied in future work.

\input{Figures/tempRhoProf.tex}

\input{Figures/tRhoProf.tex}

These calculations provide the density and temperature evolution with time for every zone \esbupdate{ (cell in the Lagrangian mesh) in the model}.  Figure~\ref{fig:profTrho} shows the density and temperature evolution for the zones in model M15bE1.69.
With the abundances from our core-collapse models and the temperature and density trajectories from these late-time calculations, we have the full input for our dust formation models.

\subsection{Dust Formation}

 During the expansion, the ejecta material cools to conditions where the gas-phase pressure-temperature (p-T) state becomes thermodynamically \textit{metastable}, and a phase transition is energetically favorable \citet{kashchiev2000nucleation}. However, the material will last in a metastable state for an extended period due to a kinetic energy barrier spanning the transient phase space. This tension is resolved through the mechanism of \textit{nucleation}; molecules in the new phase may grow by Boltzmannic attachment and eventually form large \textit{critical clusters} that are locally truly stable and provide a seed for spontaneous growth \citet{vehkamaki2006classical}.
 
 The formation of a molecular cluster of size $n$ ($n$-mer) results to a decrease in free-energy but introduces an interface between the phases that requires excess free-energy to maintain (\textit{surface tension}). Thus, the \textit{driving force} of nucleation is the difference in free-energy
 
 \begin{equation}\label{eq:diff_free_energy}
 \Delta G(n) = -G_V(n) + G_S(n)
 \end{equation}
 
 While this formulation is straight-forward, proceeding further becomes difficult. In particular, the energy required to maintain the the interface is dependent on the chemical and geometric peculiarities of the molecular structure of the $n$-mer (\citet{Mauney_2015}). Classical Nucleation Theory (CNT) simplifies this state of affairs by assuming clusters are treated as nanoscale portions of the bulk stable phase that form through attachment of monomers, and minimize the free-energy to maintain the interface by growing as dense spheres (the \textit{capillary approximation}).


 We follow the revised formulation of CNT given in Appendix A of \citet{Nozawa_2013}. This formulation has no explicit dependence on the standard pressure $p_s$, and incorporates the integrated kinetics of chemical reactions of at the time of nucleation. Further, nucleation and growth are controlled by the abundance of the \textit{key species}. The key species of a reaction is defined as the reactant with the lowest collisional frequency. In this reformulation, nucleation is represented as the reaction
 \begin{equation}\label{eq:reaction_diagram_nucleation}
 Z_{n-1} + (X + \nu_1 \fancylet{A}_1 + \nu_2 \fancylet{A}_2 + \cdots + \nu_i \fancylet{A}_i) \longrightarrow Z_{n} + (\eta_1 \fancylet{B}_1 +\eta_2 \fancylet{B}_2 + \cdots + \eta_j \fancylet{B}_j)
 \end{equation}
 where $\nu_k$, $\eta_k$ are the reactants/product stoichiometric coefficients, and $\fancylet{A}_k$, $\fancylet{B}_k$ are the reactant/product species, and $X$ is the key species.  In determination of reaction rates, the coefficient $\nu_{k,s}$ of the key species is taken as unity, and non-key species coefficients $\nu_k$, $\eta_k$ are normalized to the key species. Details of this formulation are provided in Appendix \ref{appx:nucl}.

 \subsubsection{Is CNT Appropriate?}
 The applicability of CNT to the vapor formation of dust precursors has objections. It has been argued \citep{donn1985does} that the simplifying conditions of CNT do not hold in the environment of stellar outflows where the timescale of nucleation may be longer than the timescale of condensation and we can no longer rely on steady-state assumptions. Further, taking bulk material properties as transitive to the molecular scale is dubious at best and are likely to be wrong for the small grain sizes that form in astrophysical environments \citep{nuth1996grain}.
 
 To account for these objections, it is first noted that we use the analytic formula for the rate of steady-state nucleation that proceeds through chemical reactions at the time of formation \citet{Nozawa_2013}. As shown in that paper, when the key-species collision timescale $\tau_{coll}$ is much shorter than the supersaturation timescale $\tau_{sat}$, or $\tau_{coll} \lll \tau_{sat}$, steady-state and non-steady-state nucleation rates yield quantitatively similar results, and further that $\tau_{sat}/\tau_{coll}$ typically ranged from 10-10$^5$ in the environments of type-II SNe (see Figure 9 of \citet{Nozawa_2013}).
 
 As to the issues in CNT due to its phenomenological foundations (for instance the difficulty of applying the capillary approximation), it is a challenge to identify an approach to modeling dust that is free of these issues. The other common approach is using chemical kinetics to evolve the population of dust grains through a sequence of chemical and molecular reactions, such as the comprehensive work of Cherchneff, Sadavoy, and Sarangi \citet{sadavoy2019life, sarangi2013chemically, sarangi2018dust}. However, these methods also rely on equilibrium reaction rates to evolve their molecular systems, and these reaction rates have questionable applicability in an astrophysical environment. Determining the actual formation channel using \emph{ab-initio} methods of a single grain species is a challenging undertaking, and has only been done with precision for very few grain compositions \citet{goumans2012efficient, mauney2018formation}.
 
 Further, although not explicit in this formulation as in CNT, the same difficulty of defining molecular surfaces remains in the kinetics model of nucleation. For instance, in Sunder \citet{sluder2018molecular}, molecular collisions are calculated using cross-sections of implicitly spherical particles.
 
 It is clear, however, that calculations performed here could be improved by incorporating more non-equilibrium mechanisms. For instance, the decay heat of Ni$56$ and, more importantly, the non-thermal ionizing radiation it produces during decay, are not considered in these models. We are currently working on enhancing the capabilities of our code to account for these processes, as well as incorporate a chemical kinetics background to take full advantage of those methods \citet{mauney2016formation}.
 
\subsubsection{Implementation}

 We have implemented the model of the previous section into a python code called \textit{nuDust}\footnote{https://github.com/lanl/sndust}. \textit{nuDust} is built to use the libraries NumPy \citep{2020NumPy-Array} and SciPy \citep{2020SciPy-NMeth} for fast and accurate numerical algorithms. The numba \citep{10.1145/2833157.2833162} library is used for just-in-time (JIT) compilation of python code to produce efficient machine code, and to facilitate thread and GPU parallelization.  This code takes as input a list of chemical and nucleation reactions, an initial chemical composition, and the time-series data of a prior hydrodynamics simulation.  Lagrangian cells act as a 0-D box of vapor. We assume the vapor is composed of hot, inert \textit{monomers}, with a chemical composition taken from the initial model setup.
 
 Time-series data of the hydrodynamics of the cell is used to construct a cubic piece-wise polynomial spline \citep{akima1970new} for interpolating values of temperature and density (Fig. \ref{fig:Fig3-44-dust}). Before integration begins, the initial concentrations are modified by assuming the complete formation of the fast-forming molecules CO and SiO. 
 
  At the beginning of each time-step, temperature $T$ and density $\rho$ values are evaluated, along with their derivatives. The system of ODEs are simultaneously evolved for every species until all key species have been exhausted or the temperature of a cell falls below a threshold value where there will be no further chemical activity w.r.t. nucleation and grain growth. With $\Vec{x}$ as a vector of concentrations $c_i$ of $N$ chemical species and $\boldsymbol{K}_{j} = (K_j^0, K_j^1, K_j^2, K_j^3)$ of $M$ grain moments
  
  \begin{equation} \label{eq:ode_xvar}
      \Vec{x} = \left(\{c_i\}, \{ \boldsymbol{K}_j\} \right) \quad i = 1, \dots, N, \quad j = 1, \dots, M
  \end{equation}
  we solve the initial value problem
 
 \begin{equation} \label{eq:ode_0}
 \begin{split}
     \frac{d \Vec{x}}{dt} &= f(\Vec{x}, t) + \Vec{x}_{c} (d\rho/\rho) \\
     \Vec{x}_{c} &= \left(\{c_i\}, \dots, 0, \dots \right) \quad i = 1, \dots, N
 \end{split}
 \end{equation}
 where $f(\Vec{x},t)$ is constructed from Eqs. (\ref{eq:nucl_massconserve_simple_again}, \ref{eq:nucl_moments}, \ref{eq:nucl_graingrow}). The second term in Eq. (\ref{eq:ode_0}) accounts for the change in the concentrations of chemical species (though not the grain moments) due to the volume change of the Lagrangian cell.
 
 The LSODA (\citet{hindmarsh1983odepack}) integrator provided by SciPy is used for integrating Eq. (\ref{eq:ode_0}). This integrator uses automatic selection of non-stiff and stiff methods. The Jacobian matrix $\boldsymbol{J} = \partial f / \partial \Vec{x}$ for implicit integration are determined numerically using finite-differencing. User-provided relative ($E_{rel}$) and absolute ($E_{abs}$) error tolerances adjust the time-step so that
 
 \begin{equation} \label{eq:ode_error}
    \|e_i\| \leq \max(\Vec{x}*E_{rel}, E_{abs}).
\end{equation}


\section{Results} \label{sec:results}

To probe the dependence of dust formation on the properties of the supernova, we first constructed a large database of supernova explosion models evolved out to a minimum of 1157 days post-explosion by continuing the hydrodynamical evolution of many of the existing results obtained by \cite{fryer2018} with the simplified FR99 code. Our database encompasses 21, 30, and 21 explosion models with $\Mprog$ = (15, 20, 25) $\Msol$, respectively, for a total of 72 explosion models covering a wide range of explosion energies, $\Eexp = \es{(0.53-18.4)}{51}\ \uerg$. As a note, seven of the 30 models with $\Mprog=20\Msol$ cover a range of $\Eexp = \es{(4.3-124)}{51}\ \uerg$ and are used to help represent lobes of single-lobe asymmetric supernova and double-lobe hypernova explosions. The database of the explosion models used is given in Table \ref{tab:fryer-models}. This large suite provides a wide probe of the explosion energy parameter space that we are investigating. 

The temperature, density ($T,\rho$) trajectories from these explosion models are used as input in our dust formation code, producing a database of dust nucleation models. All of our dust models were studied out to a minimum $t=1157\ \mathrm{days}$ to provide ample time for most of the grain species modelled to nucleate and grow before the corresponding key species were fully depleted or the simulation evolved beyond a ($T,\rho$)-space that was amenable to dust nucleation and growth. This time period is relatively short compared to the evolutionary timescale of young supernova remnants and allows us to probe the growth of dust grains in CCSN ejecta prior to the reverse shocks that occur when the ejecta interacts with the ISM at the onset of the SNR stage.

We used the moment equations described in Section \ref{sec:methods} to calculate the mass of the dust species in each model as well as the average radius of the dust grains for each grain species. Table \ref{tab:dust-masses} gives the collated results for a number of modelled grain species that contributed significantly to the dust content for each explosion model available to us.  The dust grains presented in the table have been limited to carbon-, silicon-, and oxygen-based species as the Fe-group species did not produce substantial amounts of dust except for the $25\Msol$ progenitors.

The models in Table \ref{tab:dust-masses} were grouped by their explosion model designations (e.g. M15a, M20b, etc) and ordered by explosion energy within these subgroups. It is clear that, within these energy-ordered subgroups and across the three separate progenitor classes, the amount of dust produced by $t=1157\ \mathrm{days}$ depends upon the explosion energy and progenitor.  This trend is generally observed for all productive dust species.  In this section, we review these trends focusing on the distributions and growth of the dust grains.

\subsection{Distribution of Dust in the Ejecta}

Figures \ref{fig:Fig1-71-dust}, \ref{fig:Fig2-35-dust}, and \ref{fig:Fig3-44-dust} show the dust fractions for different dust species, \esbupdate{ at time $t\approx 1157$ days}, as a function of enclosed mass for models M15cE3.43, M20bE2.6, and M25aE4.73 corresponding, respectively, to $\Mprog$ = (15, 20, 25) $\Msol$ zero-age main-sequence progenitors. The energies for these three progenitors are $\Eexp=\es{(3.43,2.60,4.73)}{51}\ \uerg$. These models were selected as examples of our three progenitor masses with large dust production and similar energies.

\input{Figures/ejecta-dust-15msol}

The figures plot the distributions of only a handful of the most abundant dust grains: $\carb$, $\sili$, $\silica$, $\magnesia$, $\enstatite$, $\alumina$, and $\forsterite$ (both grain reaction variants). The plot shows the abundances of each dust species produced versus enclosed mass of the ejecta in the top panels of each figure.  \esbupdate{ We also include the abundances of the CO and SiO gas phase molecules pre-formed in our simulations as well as the abundances of the free gases available to grain nucleation in the bottom panels (all references to free-gas dominant shells can be obtained from the bottom panels of Figures \ref{fig:Fig1-71-dust}, \ref{fig:Fig2-35-dust}, and \ref{fig:Fig3-44-dust}). Additionally, the bottom panels show the abundances of free gas species at time $t\approx0$ days.  We would like to note that these results are for a strictly unmixed ejecta.} 

In the remainder of this subsection, we review the dust distributions of each progenitor in turn.  The top panel in Figure~\ref{fig:Fig1-71-dust} shows the distinct growth regions of the different dust grains.  In the hydrogen envelope, \esbupdate{ $\mathrm{M}_{\mathrm{coord}}\approx[4-11]$, the solar abundance pattern produces low abundance fractions ($\mathrm{X}\lessapprox 10^{-6}$) of a broad set of silicate and oxide grain species.}  Within the helium layer, \esbupdate{ $\mathrm{M}_{\mathrm{coord}}\approx[3.25-4]$, abundance shifts produce very small amounts of C, Si, FeS dust ($\mathrm{X}\lessapprox 10^{-6}$).  Significant amounts of dust are only produced in the central regions of the ejecta, corresponding to the carbon through silicon layers of the progenitor.  In the free carbon-dominant shell, $\mathrm{M}_{\mathrm{coord}}\approx[3-3.25]$, the free carbon fraction is high, producing abundant carbon dust. The abundance fraction of carbon dust in this region ranges from 0.1-0.4 in the top panel of Figure \ref{fig:Fig1-71-dust}. Comparing this to the bottom panel of the same figure, we see that the abundance fraction of free carbon for this same region is nearly identical, indicating a near-total conversion of free carbon into carbonaceous grains.

Moving deeper into the ejecta, we first cross a transition region between the free carbon-dominant and free oxygen-dominant shells, where the pre-formation of CO gas molecules is very high. This transition region is near fully depleted in free carbon gas and shows a strong free oxygen gas depletion curve, $\mathrm{M}_{\mathrm{coord}}\approx[2.5-3]$ with $\mathrm{X}\lessapprox [0.5-0.001]$, respectively. This C-O transition region initially contains free Mg, SiO, and Al gas abundance fractions ranging approximately within 0.01-0.03, 0.002-0.003, and 0.0001-0.0003, respectively. Subsequently, we see the modest formation of $\forsterite-$Mg, $\mathrm{X}\approx 0.01$, and limited formation of $\enstatite$, $\alumina$, and $\forsterite-$SiO, $\mathrm{X}\lessapprox 10^{-5}$.

Once we are fully in the free oxygen-dominant shell ($\mathrm{M}_{\mathrm{coord}}\approx[2-2.75]$), $\forsterite$ (both reactions combined) and $\alumina$ dominate the abundances with approximately 0.2-0.35 and 0.03-0.05 of the abundance fractions taken, respectively. There is also a spike of $\enstatite$ around $\mathrm{M}_{\mathrm{coord}}\approx2.1$ with an abundance fraction of 0.001. Interestingly, around $\mathrm{M}_{\mathrm{coord}}\approx[1.8-2.2]$, we see steep drop off of $\forsterite-$Mg abundances by several orders of magnitude before vanishing to zero at $\mathrm{M}_{\mathrm{coord}}\approx2.1$. This feature coincides with a strong increase in $\forsterite-$SiO abundances by a two orders of magnitude over the same region of mass coordinates. Moving minimally deeper into the ejecta, we arrive at the transition from free oxygen-dominant to free silicon-dominant shells that contains the highest abundance of free SiO gas molecules. Unsurprisingly, we see large abundances of this material go into $\silica$ formation comprising 70 percent of the abundance fraction.}

\input{Figures/ejecta-dust-20msol}

Figure \ref{fig:Fig2-35-dust} shows the corresponding images for the $20\Msol$ progenitor model where the dust species follows the same trends as the $15\Msol$ progenitor dust distribution. However, there are noticeable differences between the two models, with the first difference being that the distinct regions of dust growth are extended in mass coordinate due to the larger ejecta mass and corresponding progenitor composition shells. \esbupdate{ For example, the carbon rich layer, $\mathrm{M}_{\mathrm{coord}}\approx[4.75-5.75]$, spans a region $\Delta\mathrm{M}_{\mathrm{coord}}\approx1.0$ for the $20\Msol$ progenitor compared to $\Delta\mathrm{M}_{\mathrm{coord}}\approx0.5$ for the $15\Msol$ mass progenitor, resulting in a larger total carbon dust mass for this shell within the ejecta.}  \esbupdate{ In the O/Mg/Al region of the ejecta, $\mathrm{M}_{\mathrm{coord}}\approx[2.25-4.75]$, the production of $(\forsterite)_{Mg}$ dominates silicate production, followed by $(\forsterite)_{SiO}$ and $\enstatite$ production. We similar $\alumina$ production comprising the second largest abundance fraction of the dust species in this layer. It should be noted that while these regions are extended along mass coordinate in comparison to Figure \ref{fig:Fig1-71-dust}, the abundances of each of $\alumina$ and the silicates shown are reduced by about one order of magnitude each for most of the region of the ejecta.} We also observe the same silica abundance spikes \esbupdate{ (up to 50 percent of the abundance fraction)} with an additional layer of pure silicon dust \esbupdate{ (peaking at 60 percent of the abundance fraction)} at the oxygen-silicon interface \esbupdate{ occurring around $\mathrm{M}_{\mathrm{coord}}\approx[1.8-2.2]$}. While it is not shown here, it should be noted that for models with explosion energies 2.75 $\lessapprox$ Eexp $\lessapprox$ 5 foe, the pure silicon dust spike does not occur due to insufficient free-Si remaining after SiO gas phase production. This is possibly related to the dependence of key nucleosynthetic yields on the explosion energetics.

\input{Figures/ejecta-dust-25msol}

Finally, Figure \ref{fig:Fig3-44-dust} shows the same plots but now for a $25\Msol$ progenitor. This high mass progenitor model more resembles the lowest mass progenitor model given in Figure \ref{fig:Fig1-71-dust}, with the distinct regions of dust production occurring in \esbupdate{ extended mass coordinate shells because of the larger ejecta mass. These regions contain $\approx$3 times as much mass for the carbon, O/Mg and O/Si layers when compared to the $15\Msol$ progenitor in Figure \ref{fig:Fig1-71-dust}.} We see all of the same features as noted from before. \esbupdate{ One comment about the overall abundance fractions for the dominant dust species in the O/Mg layer, we see that only $(\forsterite)_{SiO}$ and $\alumina$ have the same drop in typical abundance by about one order of magnitude as seen also in the $20\Msol$ progenitor of Figure \ref{fig:Fig2-35-dust}.} We also observe a large silicon dust feature in the innermost ejecta.

\subsection{Growth of Grain Mass}

With our set of models, we can also study the growth of dust in terms of mass and average radius as a function of time. Various species of dust grains will form at different regions of ($\mathrm{T},\mathrm{\rho}$)-space which will impact the time at which these grains can be observed at post-explosion. For this analysis, we continue to partition our results by the progenitor masses of the explosion models.

\esbupdate{ We first give the results of dust production as a function of time for a select number of grain species and explosion models for each progenitor class in our database in Figure \ref{fig:dustmass-growth-time-species}. We show the mass of $\carb$, $\alumina$ $\enstatite$, and $\forsterite$ in purple, cyan, olive, and red lines, respectively as it evolves over time throughout the dust simulation from 0 to 1157 days. We note that the general trend seen in all panels of Figure \ref{fig:dustmass-growth-time-species} is that as explosion energy increases, the time at which bulk grain production occurs is earlier and earlier. This trend is also generally agnostic of the grain species, indicating that this result is potentially directly tied to the explosion energetics.

\input{Figures/dust-species-mass-growth-time}

For example, looking at the top panel of the figure, with 15 $\Msol$ progenitors, we see that for the 1.86 foe model, bulk carbon growth occurs around the 800 day mark, whereas this bulk growth occurs around 700 and 650 days for the 2.6 and 3.42 foe models, respectively. For the middle panel with 20 $\Msol$ progenitors, the effect is even more pronounced with bulk carbon growth occurring at 550, 650, and 800 days for 2.6, 1.47, and 0.85 foe explosions. Going to the bottom panel with the largest 25 $\Msol$ progenitors and the dust production bulk carbon production occurs even earlier at around 350, 375, and 500 days for 4.73, 2.78, and 0.99 foe explosions. This uncovers another trend in that progenitor mass is correlated with the time at which bulk dust production occurs, not only for carbon grains, but for the other grain species presented in Figure \ref{fig:dustmass-growth-time-species}. That is, one requires less energetic ejecta to obtain earlier bulk dust production for various grain species as one increases the progenitor mass of the star.

\input{Figures/total-dustmass-growth-time-all-models}

Figure \ref{fig:total-dustmass-w-time}, reports the total dust mass for graphite, silicates and all species grouped together as a function of time for our suite of models for up to 1157 days of dust production post-explosion.  The total dust mass for each model is plotted along with the left-hand column with explosion energy color-coded with the given color-bar to the side of each row of panels. For the left-hand column of panels, the models with explosion energy less than 2 foe are given as solid lines and more energetic models are presented with dashed lines. The dust masses of carbon grains and silicate grains are given in the right-hand column, denoted by solid and dashed lines, respectively. The same explosion energy color-coding applies for the right-hand column of panels. Each row of plot panels represent the 15, 20, and 25 $\Msol$ progenitor models of each explosion for the top, middle, and bottom rows of panels, respectively. 
We first look to the left-hand column of panels in Figure \ref{fig:total-dustmass-w-time} for each set of progenitor star models. For the 15 and 20 $\Msol$ progenitor sets, we see the same trend that was elucidated in Figure \ref{fig:dustmass-growth-time-species} where explosion energy will affect the time at which bulk dust production will occur with very few exceptions. In the left-top panel, we see that earliest bulk production occurs around 625 days for a model with explosion energy around 8-10 foe and the latest bulk production occurs around 1000 days or later for all models with explosion energies less than 2 foe. The delay time for bulk dust production spans more than 500 from earliest producer to latest producer over an explosion energy range of 0.5-11 foe for these 15 $\Msol$ progenitors. Looking to the left-middle panel with 20 $\Msol$ progenitors, we see for an explosion energy range of $\approx$1-125 foe, the delay time in total dust production spans a range of about 60 days to 1000 days with the delay time increasing with decreasing explosion energy. Both the left-top and left-middle panels show a strong correlation between explosion energy and delay time post-explosion for the bulk production of dust grains. At 1157 days, the distribution of total dust mass ranges within 0.0001-0.2 $\Msol$, with the majority of these 15$\Msol$ progenitor models having total dust masses of at least 0.02 $\Msol$.

A trend more readily seen among the 20 $\Msol$ progenitors a longer time for which the initial bulk production occurs (left-middle panel of Figure \ref{fig:total-dustmass-w-time}). We see that for the explosions stronger than 2.0 foe denoted by the dashed lines in the left-middle panel, the initial bulk dust production occurs very rapidly on the order of days to perhaps a few weeks, culminating in total dust masses of 0.01-0.1 $\Msol$ of dust. For the weaker explosions, this process is noticeably slower, occurring in two stages, the first stage lasting on the order of 10-100 days with this extended first phase growing longer with decreasing explosion energy. The second phase of bulk dust production for these weaker explosion models is relatively short lived and culminates in dust masses for individual models of 0.02 to 0.1 $\Msol$. Inspecting the dust mass curves once they begin to flatten also reveals that the largest producers of dust coincide with the 20 $\Msol$ progenitors with explosion energies of 5-75 foe. More powerful explosions ultimately produce less dust, similarly to the much weaker explosions, by the 1157 day post-explosion mark. At 1157 days, these models have a total dust mass evenly distributed within the range of 0.2-0.2 $\Msol$, similarly to the series of 15$\Msol$ progenitor models. There appears to be no correlation between total dust mass and explosion energy for models that have mostly stopped dust production.

From the right-hand panels of Figure \ref{fig:total-dustmass-w-time}, we can compare the production of carbon and silicate dust species.  The carbon dust, produced further out in the star, is synthesized prior to the total sum of silicate species. In the 20\,M$_\odot$ progenitor, the time lag between bulk carbon and bulk silicate production increases with a decrease in explosion energy, the time lag between carbon and silicate production is about 50-100 days for the strongest explosions ($\approx$90-125 foe), decreases to 150-200 days or more for less energetic explosions. This trend is generally seen with the 15 solar mass models (but at later times), with silicate production generally lagging behind carbon production by about 150-200 days for most models. The range of explosion energies covered by these models is not as substantial as the 20 $\Msol$ progenitor set, but still elucidates the length of carbon-silicate production delay time correlation with explosion energy. Again, these trends are not as strong in the the 25 solar mass progenitor models, except for the time lag trend occurring between bulk carbon and bulk silicate dust production. Another common feature is that silicates ultimately produce more dust by mass than the carbon species.

Finally, we come to the 25 $\Msol$ progenitor set. Looking at the left-bottom panel, the previously stated correlation between bulk dust production and explosion energy is much less pronounced. There seems to be a tendency for middle range explosions ($\approx$5-12) to produce bulk dust around the same time ($t\approx200$), or even sooner by more than 100 days, as the strongest explosions ($\approx$12-18 foe). Furthermore, the there is still a production delay time between carbon and silicate dust species that generally increases with decreasing explosion energy, with the shortest delay times being as small as $\approx$10-20 days for highly energetic models and as large as 200 days for the weakest explosions. This series of progenitor models, however, produces more total dust than the two lower mass progenitor sets, with the mass of total dust ranging from 0.06-0.7 $\Msol$, with the majority of these models having total dust masses of at least 0.3 $\Msol$

However, it should be reiterated that the ejecta used for each model is unmixed. As the ejecta evolves in a Sedov-like trajectory, the carbon layer sits on the outermost edge of the bulk ejecta, thus it will be the first layer to sufficiently cool by adiabatic expansion for grain nucleation to occur in earnest. With the bulk of free O, Mg, Al, and Si existing deeper in the ejecta, it will remain denser and hotter for longer than the carbon layer and will not be able to nucleate until later times. A mixed ejecta may change the timing of bulk formation for different species groups and would need to investigated in future studies. These results can be seen more as an upper bound of sorts on dust production and ejecta tracing.}

\subsection{Growth of Average Grain Radius}

Another aspect of dust grains to analyze is the grain size and is especially important when considering dust survival/destruction.  As the supernova ejecta evolves into the interstellar medium, it decelerates, producing a reverse shock that can heat the dust and destroy it. Figure \ref{fig:dustradius-growth-time-species} shows average grain sizes as a function of time for the same set of models as seen in Figure \ref{fig:dustmass-growth-time-species}.  

\input{Figures/dust-species-radius-growth-time}

\esbupdate{ First looking at the top panel of Figure \ref{fig:dustradius-growth-time-species}, we can see that for carbon grains, the average grain radius is $r_{ave}\approx(8, 6, 5)$ microns for the $15\Msol$ progenitor explosions with energies $E_{exp}=(1.86,2.6,3.43)$ foe, respectively. Additionally, $alumina$ grains reach average radii of $r_{ave}\approx(2.8, 2.2, 2)$ microns for explosion energies $E_{exp}=(1.86,2.6,3.43)$ foe. The $\enstatite$ grains are still growing at 1157 days, but have all reached a minimum average radius of $\approx[3-4]$ microns, with the 2.6 foe model having a marginally larger average radius. Finally, we see that the $\forsterite$ grains (both pathways grouped together) reach average radii of $r_{ave}\approx(\-, 14, 11)$ for explosion energies $E_{exp}=(1.86,2.6,3.43)$ foe, with the third model still growing at this timestamp. A general trend seen for these dust grains, and most strongly in the carbon grains, is that the average grain radius for a given grain species increases with decreasing explosion energy.

Moving to the middle panel of Figure \ref{fig:dustradius-growth-time-species}, we can inspect the average dust grain radii for the select species used among three $20\Msol$ progenitor models. First, the carbon grains again show substantial variation in average radii, with $r_{ave}\approx(10, 8, 5)$ microns for explosions energies $E_{exp}=(1.86,2.6,3.43)$ foe. It should be noted that the $20\Msol$ model with $E_{exp}=2.6$ foe produced carbon grains with an average radius of a $\approx$10 microns, about 20 percent larger than the $15\Msol$ progenitor of the same explosion energetics with carbon grains of radius $\approx$8 microns on average. The alumina and enstatite grains for some of the models in the middle panel are still growing at time $t=1157$ days, but we can at least inspect their average sizes at this timestamp. The enstatite grains span average radii of $r_{ave}\approx(0.8, 3, 2)$ microns for $E_{exp}=(1.86,2.6,3.43)$ foe, indicating no clear trend with explosion energy and dust grain radius. The alumina grains are the least interesting at the final timestamp as they are clustered around 2 microns in radius with the lowest energy model having slightly smaller, but faster growing grains based on the growth line slope from 1000 days to 1157 days.

Finally, we have the bottom panel of Figure \ref{fig:dustradius-growth-time-species} showing results for a select number of grains and explosion models from the 25 $\Msol$ progenitor group. As with Figure \ref{fig:dustmass-growth-time-species}, the trend for these models is not as straightforward and consistent with the two lower mass progenitor sets. However, the carbon dust carries the same trend of lower energy leading to larger average dust grains. We see that the carbon grains reach sizes of $r_{ave}\approx(9, 6, 3)$ microns for models with explosion energies, $E_{exp}=(0.99,2.78,4.73)$ foe. The average carbon grain size appears to mimic the carbon grain sizes for the top panel of 15$\Msol$ progenitor mass models for a similar span of explosion energies. Examining the forsterite grain sizes, we have $r_{ave}\approx(10, 20, 11)$ microns for $E_{exp}=(0.99,2.78,4.73)$ foe. The 2.78 foe explosion model produces forsterite dust grains that are about 50 percent larger than the 2.6 foe explosion from the top panel and 100 percent larger than the 2.6 foe model from the middle panel when examined at time $t=1157$ days. The alumina grains span average radii of $r_{ave}\approx(2, 3, 2)$ microns for $E_{exp}=(0.99,2.78,4.73)$ foe, showing an approximately consistent size for these grains when compared to the lower progenitor models, regardless of explosion energy. And finally, the enstatite grains for these models span $r_{ave}\approx(3, 5, 3.4)$ microns for $E_{exp}=(0.99,2.78,4.73)$ foe, and show similar trends in average radii with the 15$\Msol$ progenitors in the top panel.}

\input{Figures/dustmass-eexp-scatter} 
\input{Figures/dustmass-eexp-scatter-late}

\section{Conclusions} \label{sec:conclusions}

\esbupdate{ We have presented a large 1-dimensional parameter study probing the affects of the supernova explosion on the formation of dust in the resulting expanding ejecta. This work has been conducted as a first stage to understanding the survival of dust upon injection into the ISM and how the supernova explosion may influence this survival. In our results, there are a number of trends that appear within our large set of dust formation models. The most predominant trends appear to be most correlated to the gas dynamical evolution of the expanding ejecta that is dictated by the energetics of the preceding supernova explosion. As illustrated in Figures \ref{fig:dustmass-growth-time-species} and \ref{fig:total-dustmass-w-time}, time of bulk dust production, irrespective of individual grain species, is generally affected by the supernova explosion energy. That is, bulk production occurs earlier for more energetic explosions as these explosions evolve more rapidly due to higher initial kinetic velocity.

Furthermore, there is a correlation between time of bulk grain growth and the progenitor mass, where an increase in progenitor shortens the bulk production time when holding the explosion energetics constant. It is seen that bulk graphite production occurs typically 100-300 days before bulk production of alumina and forsterite, with the delay time of bulk production being even larger for enstatite for the 25 $\Msol$ models.
}

However, there is growing evidence that the supernova explosion is highly asymmetric~\citep[for a review, see][]{2007IJMPD..16..941F} and additional observations continue to support this claim~\citep{2014Natur.506..339G}.  The asymmetries in the explosion will grow as the shock moves through the star and subsequent circumstellar medium, driving strong mixing.  Future studies with a realistically mixed ejecta would be useful to determine if some of the more specific trends, such as the early carbon dust formation, is a product of an unmixed ejecta or not. This motivates the need for multidimensional studies, as well. The extent of mixing will alter the formation history of specific species, but should not substantially impact the species agnostic gas dynamical dependence on the explosion.

\esbupdate{ We can make comparisons to previous numerical studies as a first-pass code validation. The dust evolution of 12, 15, 19, and 25 $\Msol$ progenitors with $10^{51}$ ergs energetics, was modeled from 100 to 1500 days in \citep{sarangi2013chemically} and serves as a useful starting point for comparisons. Looking at their Table 4 of results for 15 $\Msol$ progenitor explosions at 1500 days post-explosion, they report dust masses of 5.6(-3), 1.1(-4), 7.8(-3), 3.9(-4), 2.3(-2), and 6.1(-4) (using their notation in $\Msol$) for forsterite, silica, alumina, pure silicon, pure carbon, and silicon carbide, respectively with a total dust mass of 0.038 $\Msol$ for the $^{56}Ni=0.075 \Msol$ case. For the $^{56}Ni=0.01 \Msol$ case, these values are 2.6(-6), 1.1(-4), 7.9(-3), 3.8(-4), 2.4(-2), 5.0(-4), for same ordering with 0.059 for the total dust mass (all in $\Msol$). In terms of energetics and progenitor mass, model M15bE0.92 from Table \ref{tab:dust-masses} compares best with dust masses of 8.84(-6), 2.00(-7), 4.21(-7), 0.0, 4.31(-2), 6.10(-12) for the same ordering of dust species with a total dust mass of 0.0431 $\Msol$. It should be reiterated that these numbers are reported at 1157 days when most non-carbonaceous species are still forming at this time for low energy models.

In general, the explosion energetics for lower energy models serves as a reasonably good parameterization of the time at which bulk dust growth occurs in our dataset, and silicates ultimately form the majority of total dust for 15 $\Msol$ progenitors with explosion energies of $\lessapprox 10^{51}$ ergs (see Figure \ref{fig:DUSTMASSTOTAL-EEXP-LATE}. For higher energy explosions for 15 $\Msol$ progenitors, it is generally seen that carbon grain production is of same order magnitude, alumina, and forsterite production are of an order magnitude larger, and all other species production are several orders or more magnitude smaller when compared to Table 4 of \cite{sarangi2013chemically}. For brevity, we will comment that our 20 and 25 $\Msol$ progenitor models share some agreement with total dust mass of the 19  and 25 $\Msol$ models given in their Tables 6 and 7, with forsterite, alumina, silica, and carbon production generally within an order of magnitude of our models at any energy that see dust production (mostly) resolved by our reported snapshot at 1157 days. It should be noted that \cite{sarangi2013chemically} used a more complex gas chemistry, but a simplified explosion modeling approach based off of \cite{Nozawa_2007}.

There is also a clear dependence of grain size of individual species on the explosion energetics, where less energetic models ultimately produce larger dust grains as seen in Figure \ref{fig:dustradius-growth-time-species}. The likely physical explanation here is that the cooling rate for weakly energetic explosions is lower than the cooling rate for highly energetic explosions. This means that for less bright supernovae, the ejecta traverses the ($\mathrm{\rho}$,T)-space amenable to dust production over a longer period of time. This is not surprising as the time-dependent integration of grain growth is linearly dependent on temperature as seen in Equation \ref{eq:nucl_graingrow} of Appendix \ref{appx:nucl}. We should ultimately see grains grow larger if they exist in a suitable T-space for a longer period of time.

However, are these grain sizes realistic? A supporting example is Figure 3 of \cite{gall2014rapid} where it is reported that only grain size distributions with grain radii larger than 0.25 microns, with a lower limit of 0.7 microns, can reproduce observed supernova extinction curves \citep{zubko_2004,brandt_2012}. In general, our grain radii span from $10^0$ to $10^1$ microns for grains comprising the majority of the dust mass fraction. This is in relatively good agreement for the early time dust formation estimates based on reported observations for SN 2010jl. An improved physics model, including more advanced gas chemistry and chemical kinetics could potentially improve this agreement. Additionally, modeling shock destruction on our grain distributions 

We would also like to make a few comments about the 25 solar mass progenitor models. The explosion data indicates that the reverse shock that occurs within the deepest layers of the ejecta at early times ($t\lessapprox 60\ days$) is weak and may not sufficiently reheat the innermost zones of the star, stalling near the proto-neutron star for some of these models. The inner layers between the stalled early reverse shock and the silicon layer of the ejecta cools at a similar rate to the outer layers of the ejecta, allowing for early dust formation deep in the ejecta. While we have not directly included radioactive heating primarily due to the $^{56}$Ni $\rightarrow$ $^{56}$Co $\rightarrow$ $^{56}$Fe decay chain, this heating source may ameliorate the early dust formation in the silicon layer of the ejecta. While this heating would affect the timing of dust formation for every progenitor set, it should not affect the ultimate result as these models still cool at a sufficiently high rate to form dust within the first decade.

While not presented here, a series of simulations were performed modeling the radioactive decay chains of all unstable isotopes in the ejecta without heating or dust formation enabled. The results of these decay simulations indicated that the primary sources of dust in our models, silicate, carbonaceous, alumina, and (to a lesser extent) other oxide grains, will generally be unaffected by changes in elemental abundances. One point to raise is the production of Fe-group grains will be affected, primarily due to radioactive nickel and cobalt decaying into stable iron isotopes, increasing the abundance of free iron gas in regions with low concentrations of oxygen. The growth of iron and iron sulfide grains will likely be enhanced in these regions.}

\subsection{Dependence of Dust Yields on Explosion Energies and Resulting Nucleosynthesis}

An interesting feature of our results is the dependence of final dust abundances on the energetics and resulting nucleosynthesis of the explosion. In Figure \ref{fig:DUSTMASSTOTAL-EEXP}, we see that the total dust mass of all models is dominated by silicate dust formation, followed by carbide dust, and then oxide dust. One interesting feature is the modest parabolic shape of the $20\Msol$ data. It appears that the dust formation of silicates peaks around explosion energies of $4-6$ foe and decreases as explosion energy further increases to extremes. This same trend is witnessed in Figure \ref{fig:DUSTMASSTOTAL-EEXP-LATE} where we have most models evolved out to various later times (typically 5-15 years), such that dust formation has halted for all or at least most of the predominate species (carbides, oxides, silicates and iron sulfide). We see that the carbon dust trend is nearly flat with explosion energy given sufficient evolution time, indicating early and efficient graphite production. We still see the parabolic peak in the silicate mass for the $20\Msol$ progenitor models. There is also substantial formation of $FeS$ for low energy models in the $15\Msol$ progenitor set. The substantial production of $FeS$ is also seen in most models for the $20$ and $25\Msol$ progenitor sets, except for the $20\Msol$ models with $\Eexp<2$ foe. Another trend seen in comparing Figures \ref{fig:DUSTMASSTOTAL-EEXP} and \ref{fig:DUSTMASSTOTAL-EEXP-LATE} is that the affects of lower explosion energy on the rate of dust production weakens with an increase in progenitor mass. 

Coming back to the silicate mass peak in the $20\Msol$ progenitor set, one explanation for this result is that the more energetic explosions will burn carbon into the constituents required for the nucleation of silicates. Not only does this remove carbon from the region of the ejecta that would otherwise be tied up in CO gas, but increases the capacity for SiO to form, a common key species of silicate nucleation, in the ejecta. This in turn allows for the increased production of silicate grain species without dramatically affecting carbon production. One important aspect to note is that there is a decrease in dust production for exceptionally energetic models. The energetics for these models is likely sufficient to further burn silicate constituents into heavier elements, such as Fe-group species, that form dust later and less efficiently. The ejecta also expands and cools the most rapidly for these particular models and they do not stay in a ($\rho,T$)-space amenable to dust production for very long, reducing the overall efficiency of the dust yields. Seen throughout the unmixed ejecta of all of our $20\Msol$ models, explosions with energetics $\Eexp \lessapprox 2.75$ foe and $\Eexp \gtrapprox 5.00$ foe have less silicon gas deep in the ejecta. The silicon gas abundance is sufficiently lower than the oxygen abundance that the silicon is entirely bound up in SiO gas. While the free silicon is dramatically reduced, it does allow for a small increase in the production of SiO dependent dust species. 

The energetics of a supernova explosion sensitively impacts the resulting nucleosynthesis, affecting the isotopic yields. This sensitivity of nucleosynthetic yields should be encoded in the dust yields of the ejecta and is generally what is observed in our database of models. As discussed in Section \ref{sec:results}, there is a peak in the $20\Msol$ silicate mass and appears to be directly related to the final yields of intermediate mass elements that are constituents of silicate species. Thus it can be concluded that the final dust yields and composition of a given explosion are dependent on the energetics. This opens up an avenue for observations as a measure of dust yields could serve to probe the nucleosynthesis of post-explosion CCSNe and test the viability of different explosion engines.

\subsection{Applications}

Beyond studying the dust production as a function of energy, our broad set of explosion energies can be combined to study asymmetric explosions.  Here we approximate the asymmetries by assuming that an asymmetric explosion can be represented by the sum of fractions of 1-dimensional explosions at different energies.  For example, hypernovae (HNe) can be represented by a portion of the ejecta represented by a strong explosion (along the jet axis) with the rest at a normal explosion energy.  In this section, we apply our multi-dimensional models to a range of explosion scenarios well-studied supernovae and supernova remnants to hypernovae.

\input{Tables/casA_1987A_HNe}

For a first application, we review observations of supernova 1987A arguing that this supernova was not spherical. The redshifted gamma-ray and iron line features~\citep{Hungerford2005ApJ...635..487H} are best fit by an explosion that has a single outflow that is much stronger than the rest of the ejecta.  By studying a range of stellar masses and combined components, we can study the expected variations in dust production for an asymmetric supernova (see Table \ref{tab:casA_1987A}).  Although the explosion energy is nearly the same for each of these models, varying the different component energies can vary the dust production by nearly a factor of 2.  In our preliminary study, the best-fitting model is our more extreme-energy 20\,M$_\odot$ model.

The Cassiopeia A remnant is more complex where observations of the innermost ejecta show multiple lobes~\citep{2014Natur.506..339G}. We mimic it by considering a series of explosive features covering a range of escape velocities from 0.5-5 times the symmetric velocity corresponding to a range of ejecta energies of 0.25-25 foe from our database.  Our set of explosion energies and the dust production for our simple Cassiopeia A model is shown in Table \ref{tab:casA_1987A}.\input{Figures/dustvm}\esbupdate{It can be seen that our simulated results are in close agreement to \cite{Biscaro2016} for total, carbonaceous, and forsterite dust masses by at most a factor of two with alumina production greater by almost a factor of ten for our results. However, our results are reported at a timestamp of nearly 3000 days with a simplified general physics implementation. The reported observational results of \citep{Priestley2019MNRAS.485..440P, delooze_casa_2019} are greater by 1-2 orders of magnitude.} The effect of hypernova explosions is much greater.  Here we assume bimodal explosions where we have a very strong component along the axis.  These strong explosions produce very different dust signatures than normal supernovae (Table \ref{tab:casA_1987A}).

These studies also allow us to  predict the evolution of dust production at high redshift.  There is evidence that the critical proto-stellar cloud mass increases at lower metallicities~\citep[for a review, see][]{2013RPPh...76k2901B,2020AJ....160...78R}, causing the initial mass function to flatten out.  At the low metallicities expected at high redshift, these massive stars will produce more pair instability and hypernovae.  Hypernovae, energetic and asymmetric supernova explosions~\citep{1999AstHe..92...87I}, are believed to be produced by rotating collapsing stars forming black holes.  The subsequent accretion disk produces a strong asymmetric explosions.  Not only do these explosions produce nucleosynthetic yields that are different than those of normal supernovae, but as we see from our models, they also produce different dust signatures.  If the IMF flattens out at high redshift, these hypernovae could dominate the number of explosions from massive stars, and observations of the dust at high redshift could constrain the amount of flattening in the initial mass function.  

The dust production depends both on the stellar mass and explosion energy and the flattening of the initial mass function can alter both of these.  Using the models for IMF evolution from Fryer et al. (in preparation) and our dust yields, we are able to estimate the dust production with decreasing metallicity or increasing redshift. Figure~\ref{fig:dustvm} shows the variation in the dust production with decreasing metallicity for a variety of evolutionary models for the initial mass function.  The increase of massive stars and hypernovae increases the production of dust.  Comparing these models observations~\citep{2020A&A...641A.168N} could be used to help constrain the initial mass function.  The pair instability fraction will also increase at low metallicity.  We have not included the production from these explosions in this study.

\bibliography{main}
\appendix
\input appendix

\end{document}

%% file: Figures/abundances.tex

\begin{figure*}[t!]
        \centering
        \includegraphics[scale=0.68]{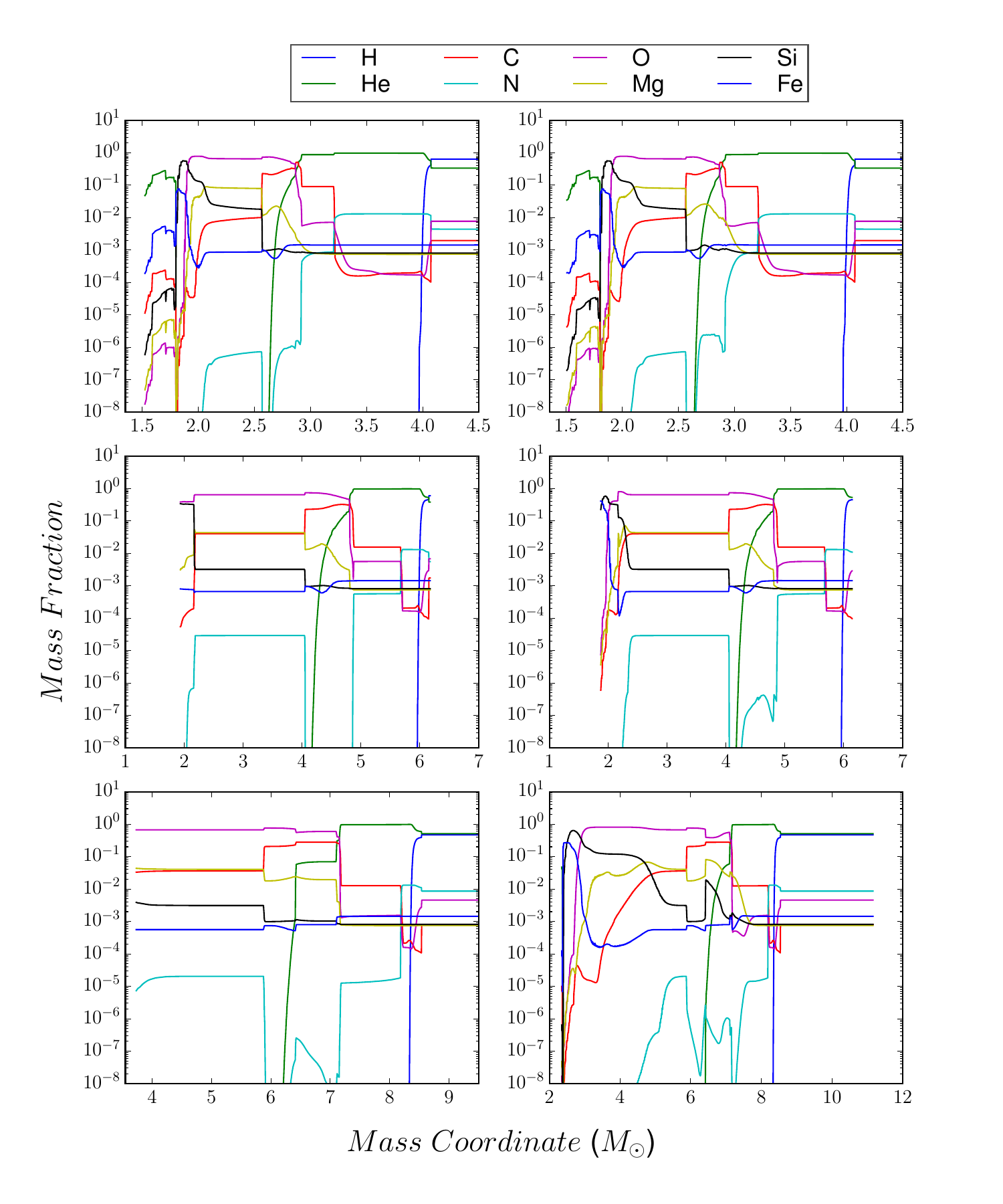}
        \caption{Above are plotted the abundance profiles for elements important in the formation of our selected grain species \esbupdate{ from dataset produced by \cite{fryer2018} using FR99}. The top row is two 15 $M_{\odot}$ progenitor models (L: 1.69 Foe, R: 3.43 Foe explosion energies), the middle row is two 20 $M_{\odot}$ models (L: 1.39 Foe, R: 5.9 Foe), and the bottom row is two 25 $M_{\odot}$ models (L: 1.57 Foe, R: 14.8 Foe). Models with the same progenitor mass have similar initial abundance profiles. However, with varying explosion energies, the distribution of nitrogen and magnesium vary the largest. With very high differences in explosion energy, the higher energy model has less uniform structure as seen in the 25 $M_{\odot}$ models. The horizontal lines in the outer regions of the profiles are due to a stitched on stellar wind.}
        \label{fig:abund}
\end{figure*}

%% file: Figures/tempRhoProf.tex

\begin{figure*}[t!]
        \centering
        \includegraphics[scale=0.68]{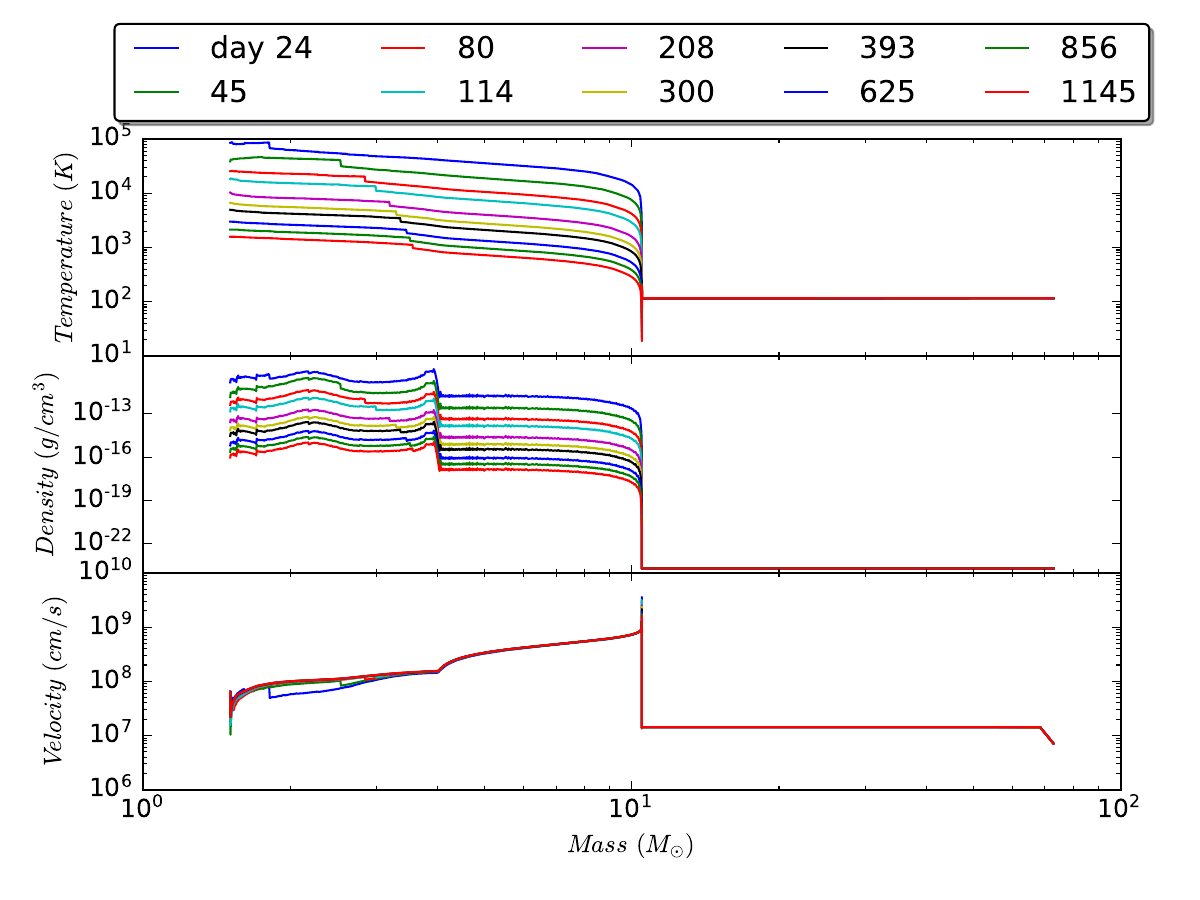}
        \caption{Top: temperature vs mass coordinate profiles for model M15bE1.69 at 24-1145 days after explosion. Middle: density vs mass coordinate profiles for the same model after explosion. Bottom: the velocity profile for the same model. An outward propagating shock can be seen at about $0.22-1 M_{\odot}$ where the temperature and density drops off as you move out in the ejecta. The shock is most prominent in the temperature plot. \esbupdate{ The sudden drop off at about $10 M_{\odot}$ indicates the interface between ejecta and the stellar wind.}}
        \label{fig:profTrhoVel}
\end{figure*}

%% file: Figures/tRhoProf.tex

\begin{figure*}[t!]
        \centering
        \includegraphics[scale=0.68]{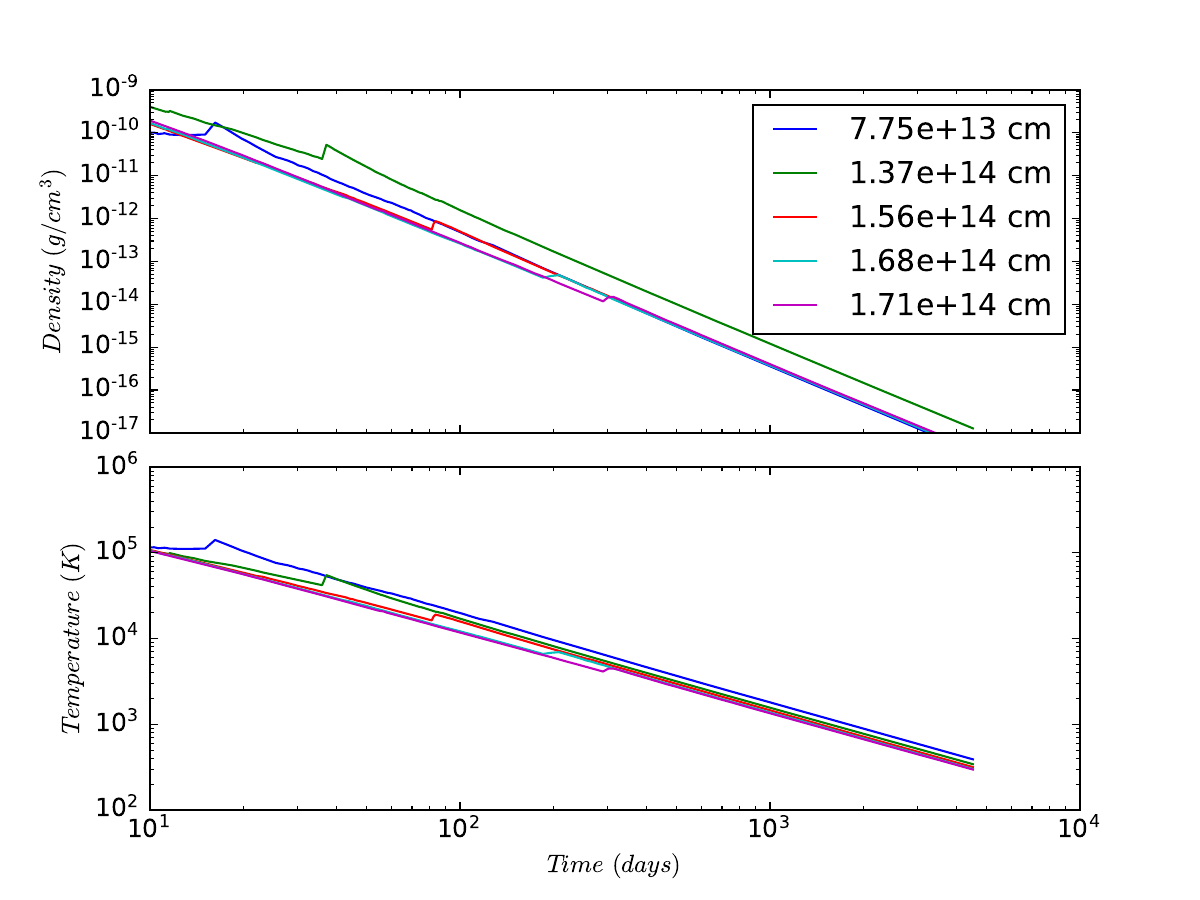}
        \caption{Top: The density as a function of time for various cells at different initial depths in the progenitor model M15bE1.69. Bottom: The temperature profile as a function of time for the same model and cells. A shock can be seen in all plots at about 20-300 days. It is characterized by an upwards almost vertical jump.}
        \label{fig:profTrho}
\end{figure*}

%% file: Figures/ejecta-dust-15msol.tex

\begin{figure}
    \centering
    \includegraphics[scale=0.95,angle=0]{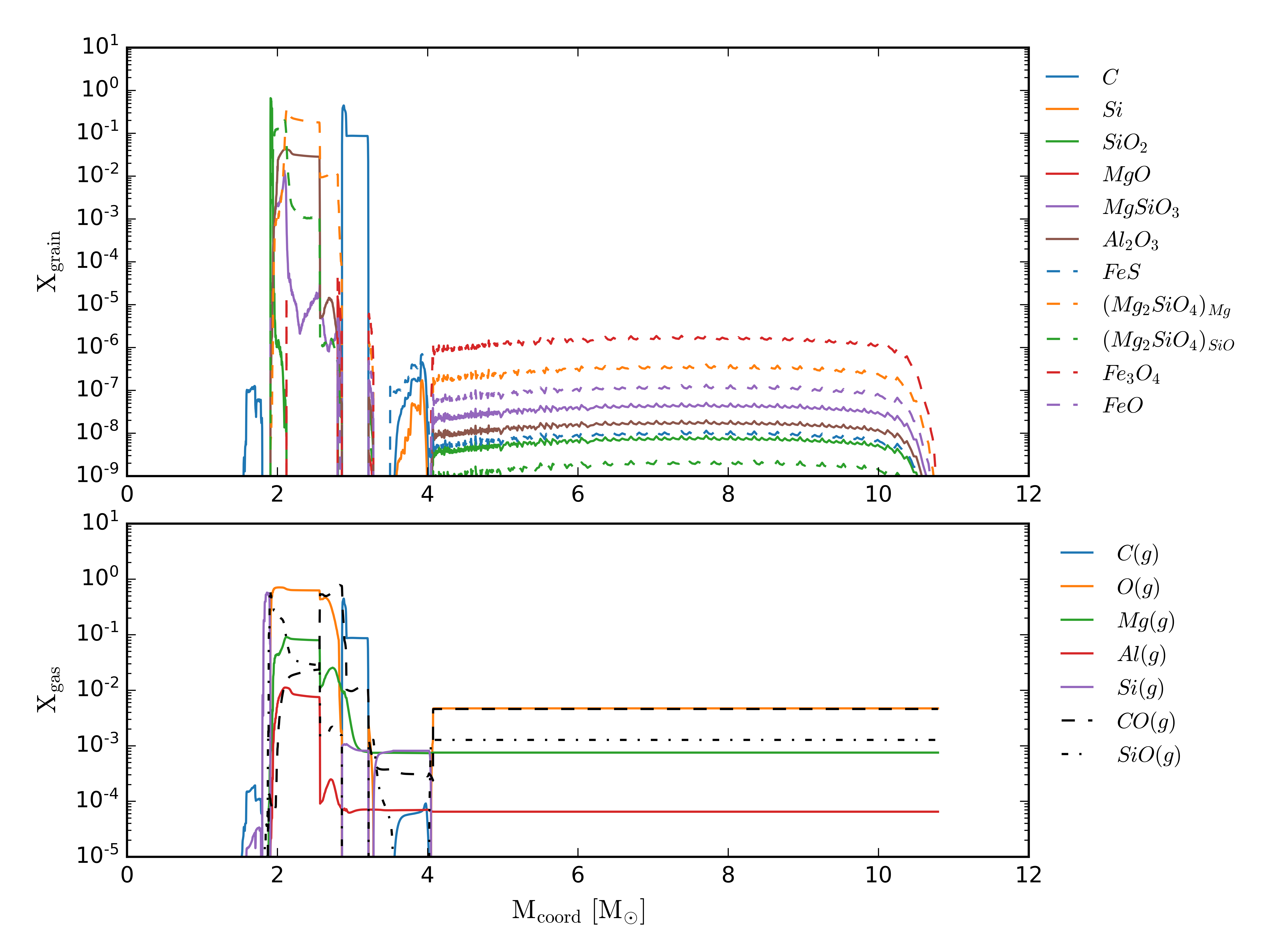}
    \caption{Top: Mass of select dust grains ($M_{grain}$) as a function of the mass coordinate of the original star given as colored lines. Bottom: Mass of gas phase elements and molecules ($M_{gas}$) as a function of mass coordinate. The gas-phase molecules CO(g) and SiO(g) are given as dashed and dotted black lines, respectively. The mass of free C(g), O(g), Mg(g), Al(g), and Si(g) are given as solid, colored lines. Both panels use data from model M15cE3.43, with $\Mprog=15\Msol$ and $E_{exp}=3.42\ \mathrm{foe}$. It should be noted, the ejecta does not model material mixing.}
    \label{fig:Fig1-71-dust}
\end{figure}

%% file: Figures/ejecta-dust-20msol.tex

\begin{figure}
    \centering
    \includegraphics[scale=0.95,angle=0]{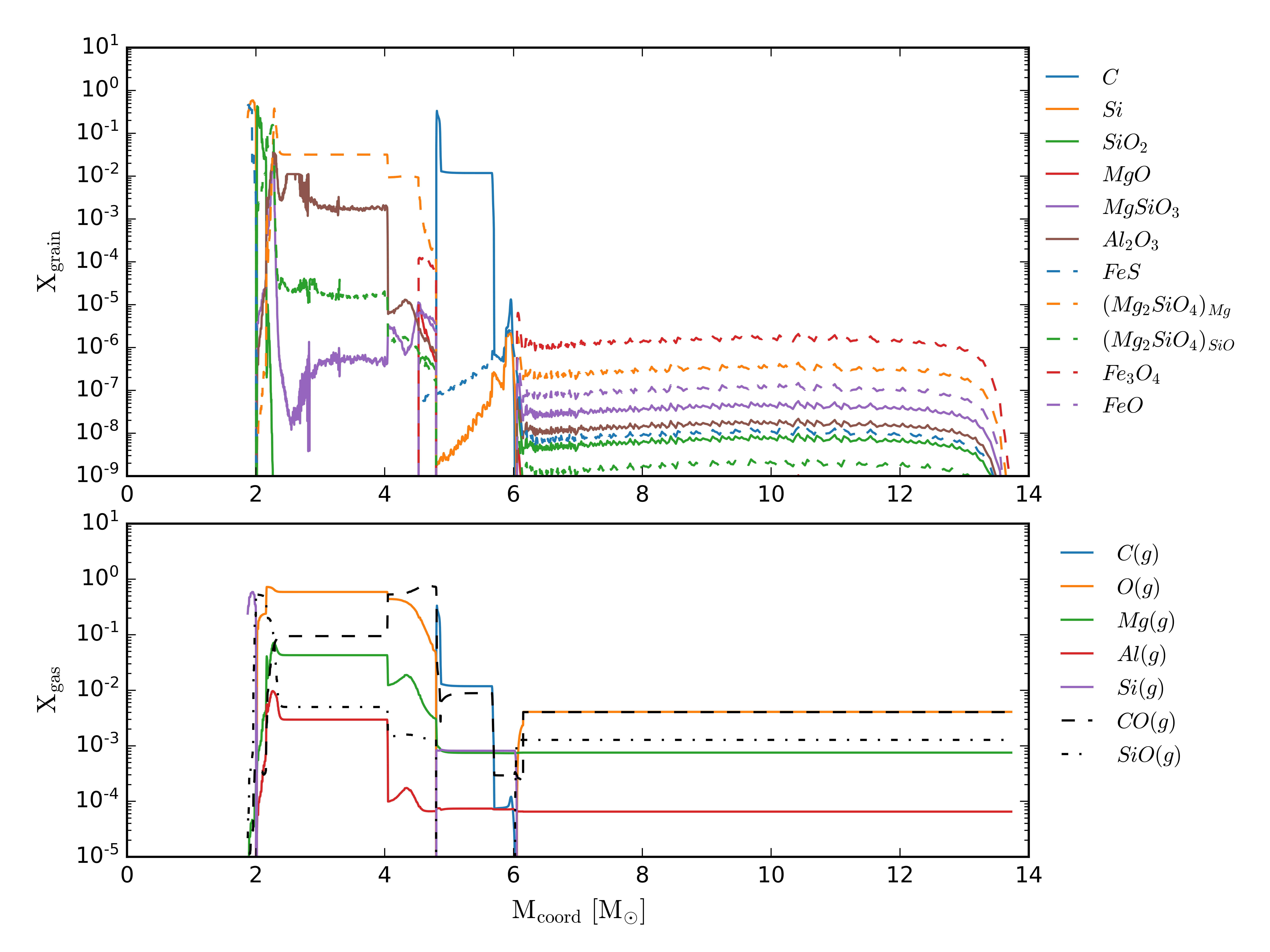}
    \caption{Top: Same as \ref{fig:Fig1-71-dust} top panel. Bottom: Same as \ref{fig:Fig1-71-dust} bottom panel. Both panels use data from model M20bE2.60, with $\Mprog=20\Msol$ and $E_{exp}=2.60\ \mathrm{foe}$. It should be noted, the ejecta does not model material mixing.}
    \label{fig:Fig2-35-dust}
\end{figure}

%% file: Figures/ejecta-dust-25msol.tex

\begin{figure}
    \centering
    \includegraphics[scale=0.95,angle=0]{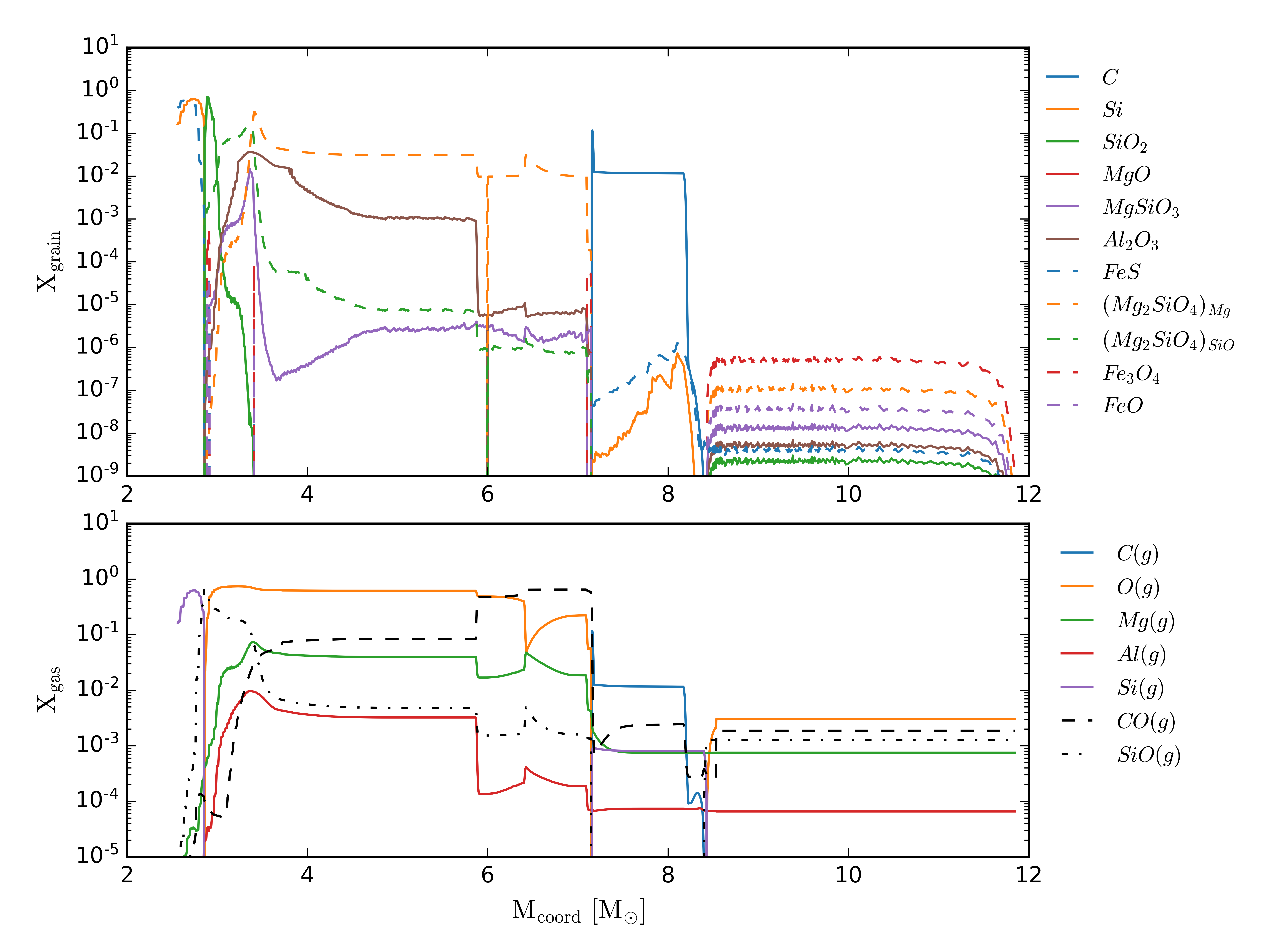}
    \caption{Top: Same as \ref{fig:Fig1-71-dust} top panel. Bottom: Same as \ref{fig:Fig1-71-dust} bottom panel. Both panels use data from model M25aE4.73, with $\Mprog=25\Msol$ and $E_{exp}=4.73\ \mathrm{foe}$. It should be noted, the ejecta does not model material mixing.}
    \label{fig:Fig3-44-dust}
\end{figure}

%% file: Figures/dust-species-mass-growth-time.tex

\begin{figure}
    \centering
    \includegraphics[scale=0.95,angle=0]{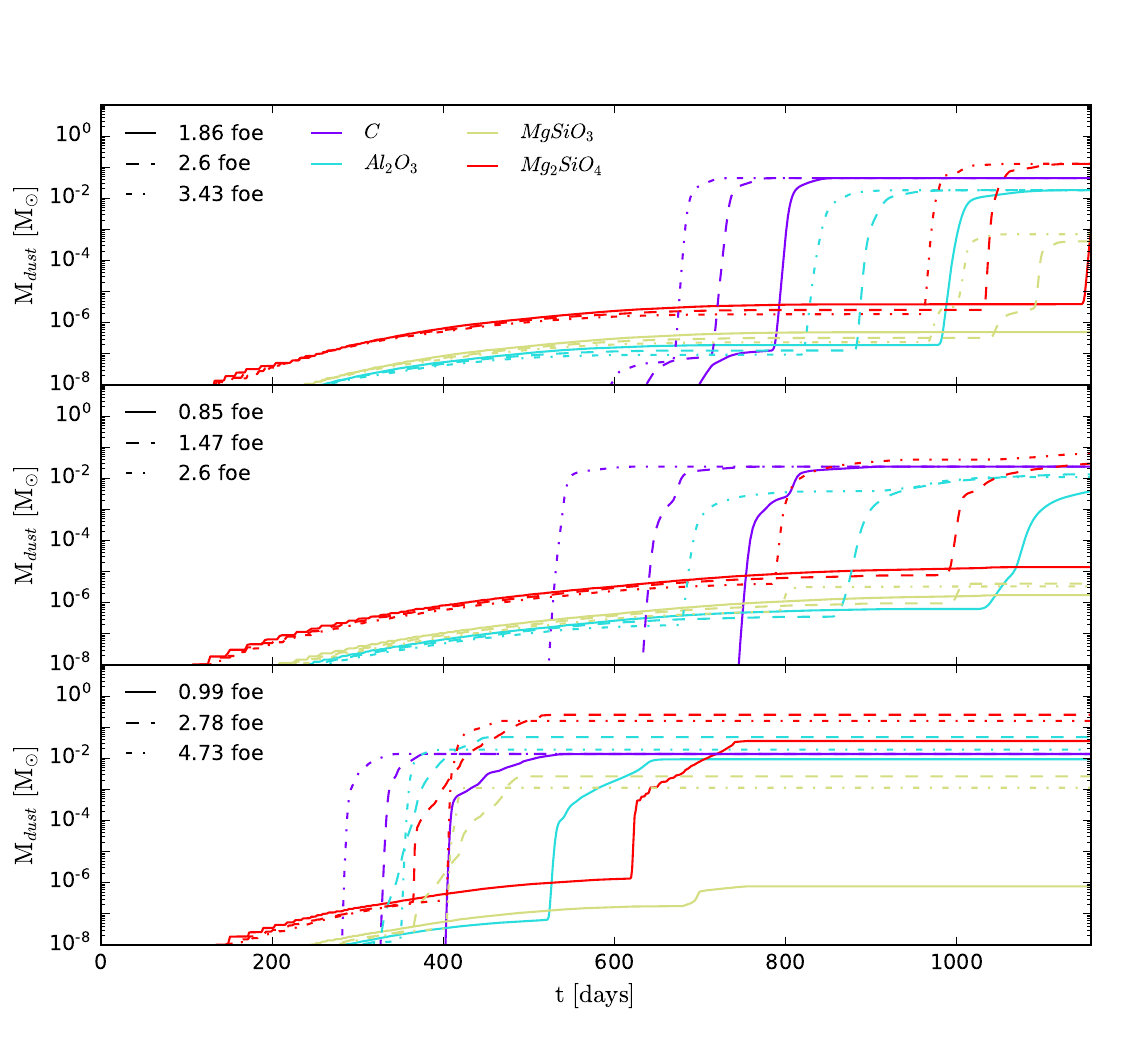}
    \caption{Top: The mass ($\Msol$) of select dust grains as a function of time after shock breakout for three CCSN models, differentiated by increasing explosion energy, with $\Mprog=15\Msol$. Grain species plotted are C(s), $\alumina$(s), $\enstatite$(s), $\forsterite$(s) as purple, cyan, gold, and red lines, respectively. The three CCSN models represented here have $\Eexp = 1.86, 2.60, 3.43\ \mathrm{foe}$ and are plotted as solid, dashed, and dash-dotted lines, respectively, for each grain species. Middle: The same as the top panel, except with three CCSN models for a $\Mprog=20\Msol$ and $\Eexp = 0.85, 1.47, 2.60\ \mathrm{foe}$ plotted with solid, dashed, and dash-dotted lines, respectively, for each grain species. Bottom: Same as the top panel, except with three CCSN models for a $\Mprog=25\Msol$ and $\Eexp = 0.99, 2.78, 4.73\ \mathrm{foe}$ plotted with solid, dashed, and dash-dotted lines, respectively, for each grain species.}
    \label{fig:dustmass-growth-time-species}
\end{figure}

%% file: Figures/total-dustmass-growth-time-all-models.tex

\begin{figure}
    \centering
    \includegraphics[scale=0.95,angle=0]{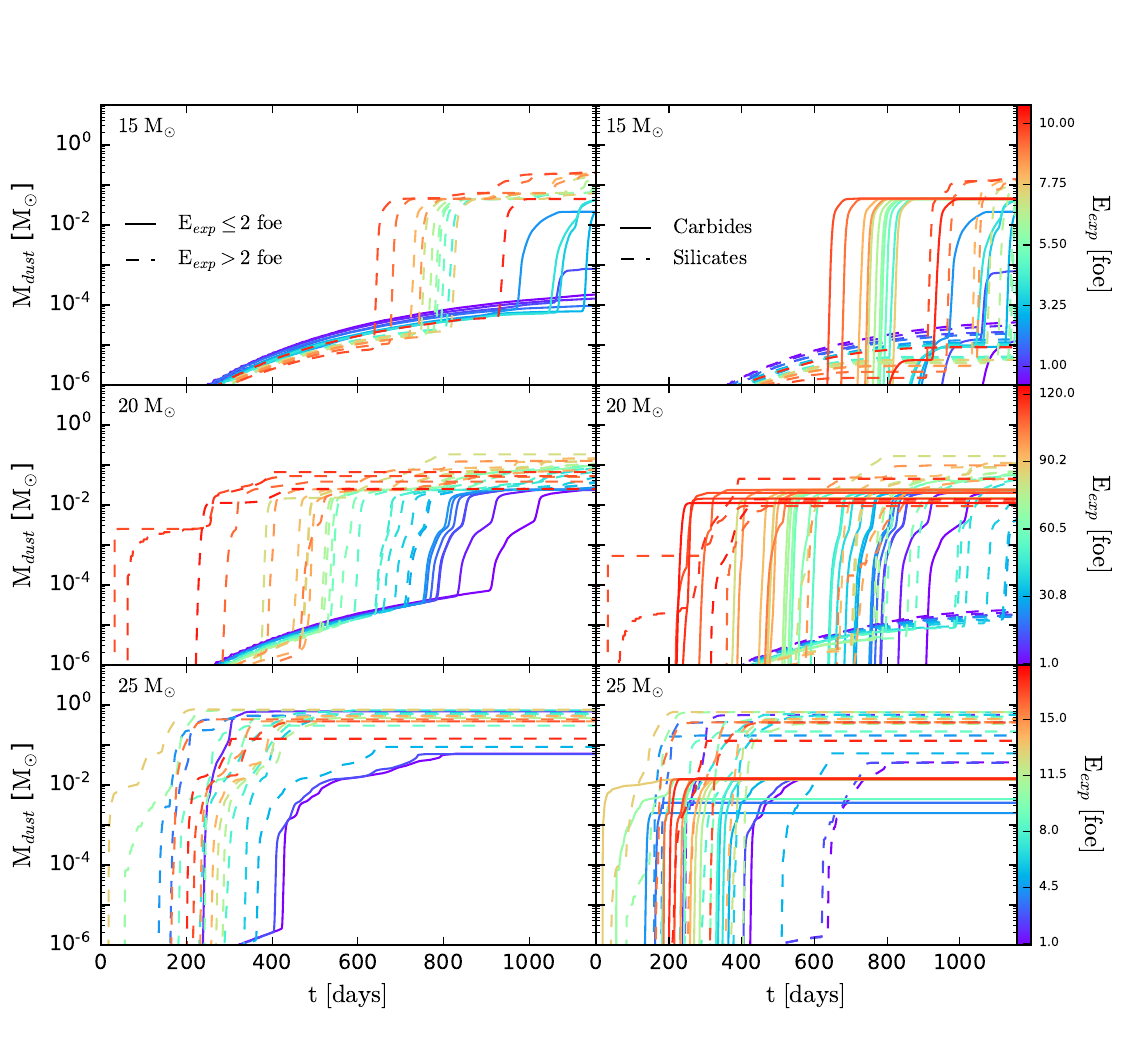}
    \caption{Top-left: The total amount of dust produced per model ($M_{dust}$), given in solar masses, as a function of time ($t$), given in days, for explosion models with a $15\Msol$ progenitor mass. The explosion energies ($E_{exp}$), in units of foe, of each model are color coated by the colorbar given adjacent to the Top-right panel. Additionally, models for $E_{exp}\leq2\ \mathrm{foe}$  and $E_{exp}>2\ \mathrm{foe}$ are given with solid and dashed lines, respectively. Top-right: Similar to the top-left panel, but now solid lines represent carbon dust grains and dashed lines represent silicate dust grains. Middle-left: Similar to top-left panel, but for $M_{prog}=20\Msol$. Middle-right: Similar to top-right panel, but for $M_{prog}=20\Msol$. Bottom-left: Similar to top-left panel, but for $M_{prog}=25\Msol$. Bottom-right: Similar to top-right panel, but for $M_{prog}=25\Msol$.}
    \label{fig:total-dustmass-w-time}
\end{figure}

%% file: Figures/dust-species-radius-growth-time.tex

\begin{figure}[h!]
    \centering
    \includegraphics[scale=0.95,angle=0]{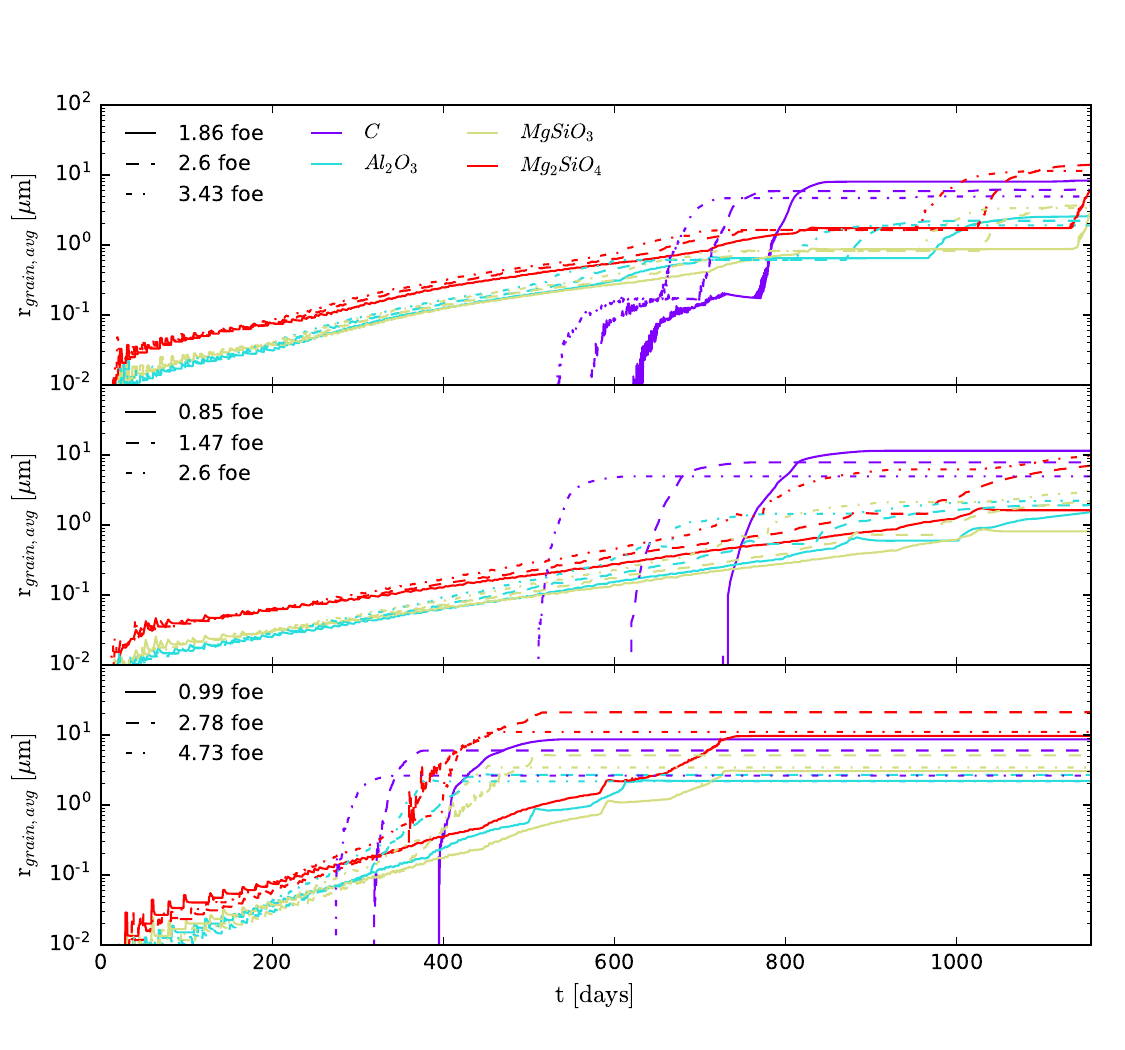}
    \caption{Top: The average radius ($\mu$m) of select dust grains as a function of time after shock breakout for three CCSN models, differentiated by increasing explosion energy, with a $\Mprog=15\Msol$. Grain species plotted are $C(s)$, $\alumina$(s), $\enstatite$(s), $\forsterite$(s) as purple, cyan, gold, and red lines, respectively. The three CCSN models represented here have $\Eexp = 1.86, 2.60, 3.43\ \mathrm{foe}$ and are plotted as solid, dashed, and dash-dotted lines, respectively, for each grain species. Middle: The same as the top panel, except with three CCSN models for $\Mprog=20\Msol$ and $\Eexp = 0.85, 1.47, 2.60\ \mathrm{foe}$ plotted with solid, dashed, and dash-dotted lines, respectively, for each grain species. Bottom: Same as the top panel, except with three CCSN models for $\Mprog=25\Msol$ and $\Eexp = 0.99, 2.78, 4.73\ \mathrm{foe}$ plotted with solid, dashed, and dash-dotted lines, respectively, for each grain species.}
    \label{fig:dustradius-growth-time-species}
\end{figure}

%% file: Figures/dustmass-eexp-scatter.tex

\begin{figure}
    \centering
    \includegraphics[scale=0.95,angle=0]{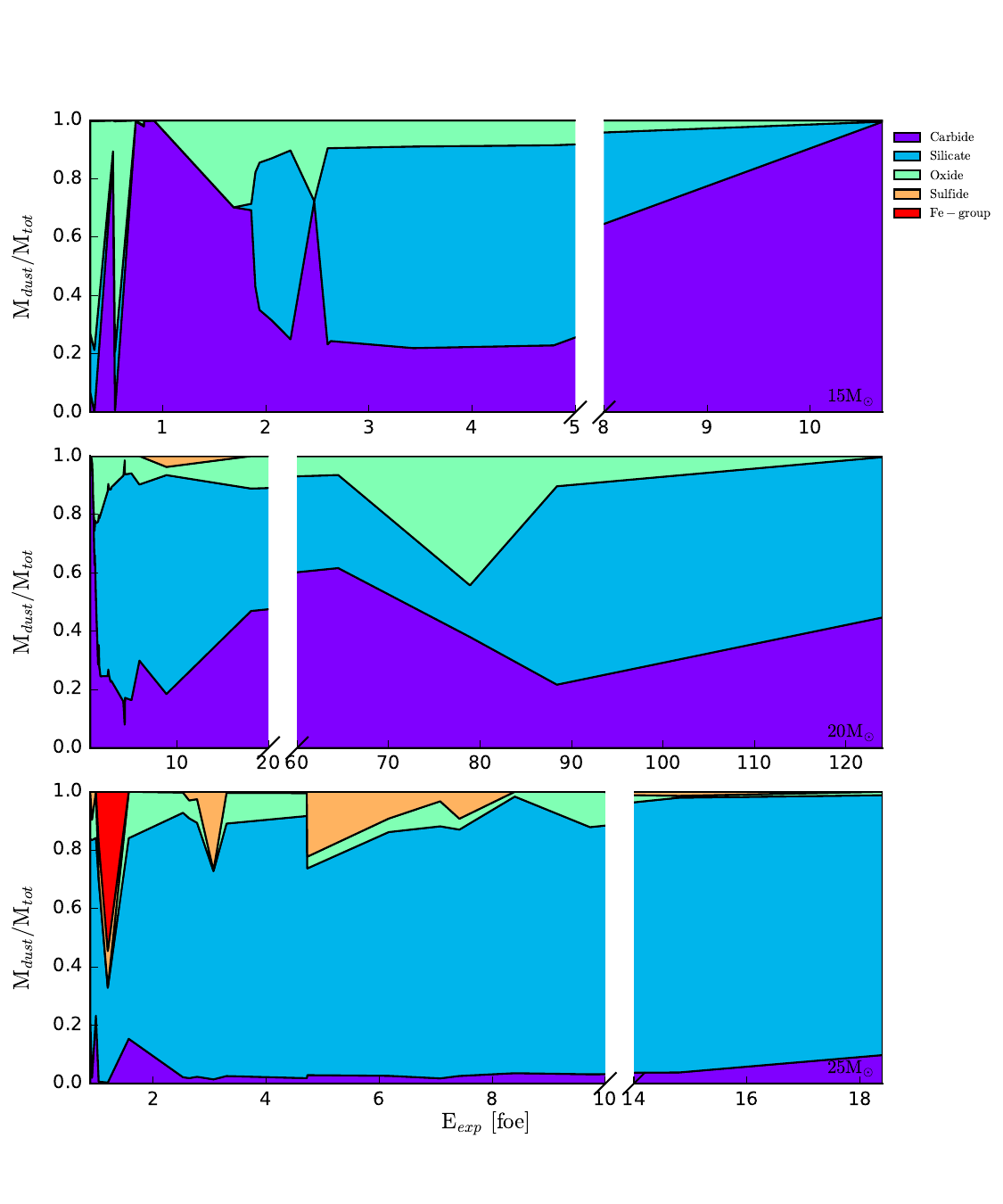}
    \caption{Top: Mass of carbonaceous, silicate, oxide, sulfide, and Fe-group dust formed normalized by the total dust for each model given as purple, blue, green, orange, and red shaded regions, respectively. The shaded regions give the portion of dust that each category takes of the total dust mass, as a function of $\Eexp$ for CCSN models with $\Mprog=15\Msol$. E.g., in the top panel from 3-4 foe, carbon, silicate, and oxide grains make up $\approx25\%$, $\approx90-25=65\%$, and $\approx100-90=10\%$ of the total dust mass, respectively. Middle: Same as the leftmost panel, except for all models with a$\Mprog=20\Msol$. Bottom: Same as the top panel, but now for only models with $\Mprog=25\Msol$. All models evolved until 1157 days.}
    \label{fig:DUSTMASSTOTAL-EEXP}
\end{figure}

%% file: Figures/dustmass-eexp-scatter-late.tex

\begin{figure}
    \centering
    \includegraphics[scale=0.95,angle=0]{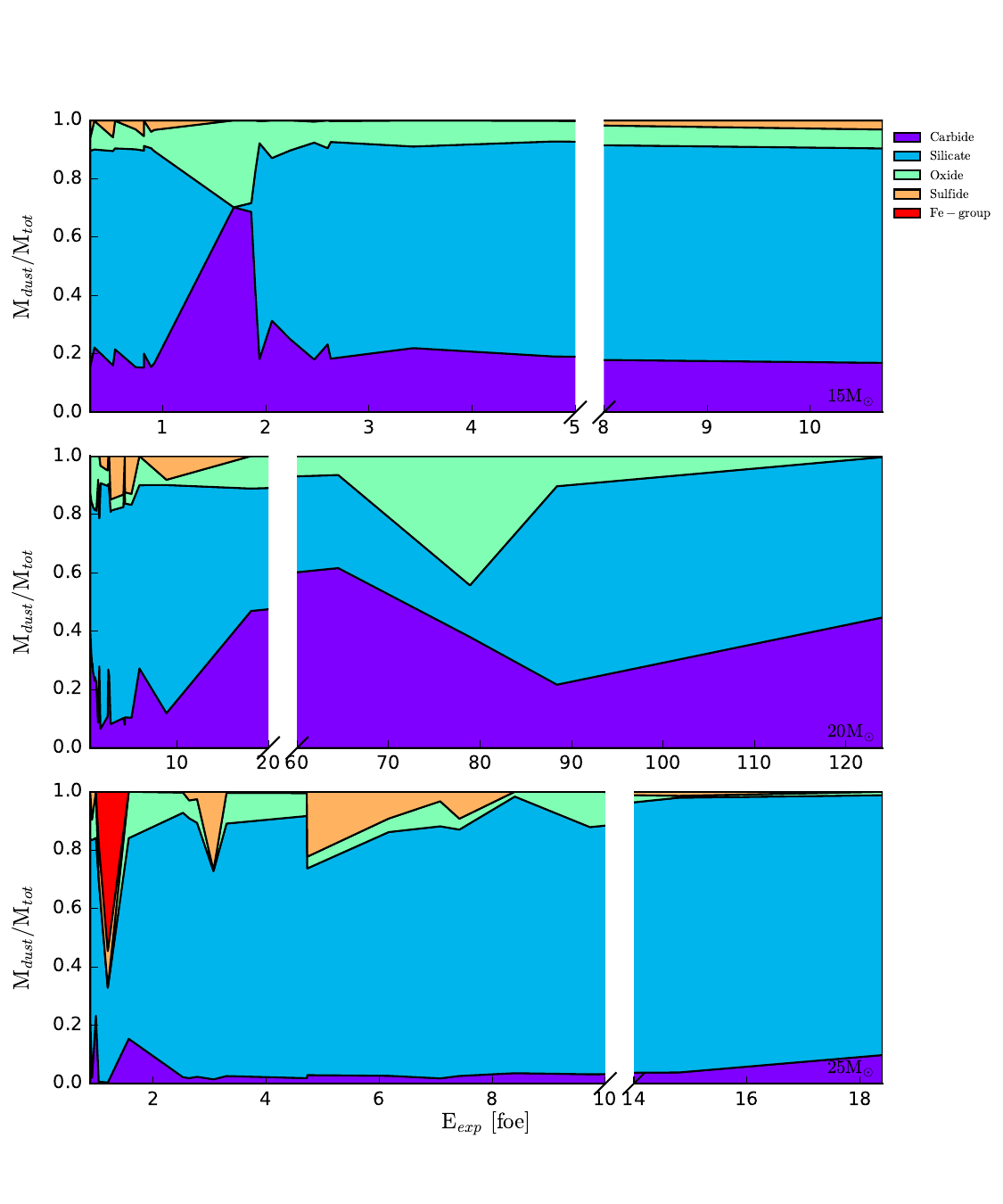}
    \caption{Same as Figure \ref{fig:DUSTMASSTOTAL-EEXP}, except models have been evolved until dust production ceases or nearly ceases, typically between 3 and 15 years.}
\label{fig:DUSTMASSTOTAL-EEXP-LATE}
\end{figure}

%% file: Tables/casA_1987A_HNe.tex
\begin{deluxetable}{c|cccccc}
\tablecaption{Comparisons to Observed Dust in Supernova Ejecta and Remnants
\label{tab:casA_1987A}}
\tablehead{\colhead{SNe} & \colhead{$\mathrm{M}_{total\ dust}$} & \colhead{$\mathrm{M}_{\mathrm{C}(s)}$} & \colhead{$\mathrm{M}_{\forsterite(s)}$} & \colhead{$\mathrm{M}_{\alumina(s)}$} & \colhead{$\mathrm{E}_{out-of-cone}$} & \colhead{$\mathrm{E}_{in-cone}$}  \\
Designation & $\mathrm{M}_{\odot}$ & $\mathrm{M}_{\odot}$ & $\mathrm{M}_{\odot}$ &
$\mathrm{M}_{\odot}$ & foe & foe}
\colnumbers
\startdata
aSN$^a$   15$\mathrm{M}_{\odot}$ & $\es{2.43}{-2}$ & $\es{6.05}{-3}$ & $\es{1.44}{-2}$ & $\es{2.00}{-3}$  & 0.52                             & 4.79                                 \\ 
aSN$^a$   20$\mathrm{M}_{\odot}$ & $\es{3.75}{-2}$ & $\es{2.36}{-2}$ & $\es{1.02}{-2}$ & $\es{1.01}{-3}$  & 0.53                             & $\frac{1}{2}$(4.33,5.03)             \\ 
aSN$^b$   15$\mathrm{M}_{\odot}$ & $\es{4.79}{-2}$ & $\es{4.32}{-2}$ & $\es{3.85}{-3}$ & $\es{5.46}{-4}$  & 0.92                             & 3.42                                 \\ 
aSN$^b$   20$\mathrm{M}_{\odot}$ & $\es{3.55}{-2}$ & $\es{2.37}{-2}$ & $\es{4.53}{-3}$ & $\es{6.65}{-3}$  & $\frac{1}{2}$(0.84,1.0)          & $\frac{2}{5}$2.85, $\frac{3}{5}4.15$ \\ 
aSN$^c$   20$\mathrm{M}_{\odot}$ & $\es{2.45}{-2}$ & $\es{2.36}{-2}$ & $\es{6.53}{-4}$ & $\es{1.68}{-4}$  & 0.53                             & 18.1                                 \\
Reference 20$\mathrm{M}_{\odot}$ & $\es{3.78}{-2}$ & $\es{2.37}{-2}$ & $\es{4.44}{-3}$ & $\es{9.61}{-3}$  & 1.00                             & -                                    \\
1987A$^{\dagger}$                & $\es{3.00}{-3}$ &  $\geq \es{2.25}{-3}$         &       $\leq \es{7.5}{-4}$      & -              & -                                & -                                    \\
\hline
HNe$^d$   25$\mathrm{M}_{\odot}$ & $\es{7.28}{-1}$ & $\es{1.37}{-2}$ & $\es{2.46}{-1}$ & $\es{5.51}{-2}$  & 4.72                             & 14.80                                \\
HNe$^e$   25$\mathrm{M}_{\odot}$ & $\es{4.71}{-1}$ & $\es{1.38}{-2}$ & $\es{1.59}{-1}$ & $\es{1.88}{-2}$  & 4.73                             & 18.40                                \\
HNe$^f$   25$\mathrm{M}_{\odot}$ & $\es{5.64}{-1}$ & $\es{1.38}{-2}$ & $\es{1.83}{-1}$ & $\es{3.73}{-2}$  & 0.92                             & 18.40                                \\
Reference 25$\mathrm{M}_{\odot}$ & $\es{4.87}{-1}$ & $\es{1.38}{-2}$ & $\es{1.85}{-1}$ & $\es{2.04}{-2}$  & $\frac{3}{4}$4.73,$\frac{1}{4}$6.17 & -                                 \\
\hline
Cas A-like 20$\mathrm{M}_{\odot}$ & $\es{3.13}{-2}$              & $\es{2.06}{-2}$ & $\es{5.89}{-3}$ & $\es{3.91}{-3}$ & 0.53, 0.81, 1.0, 1.19, 5.03, 18.1 & -                        \\
Cas A$^{\dagger\dagger}$          & $\es{7.0}{-1}$ &  $\leq \es{1.5}{-1}$          & $\es{6.0}{-1}$             & -             & -                                 & -                      \\
Cas A$^{\ddagger}$                & $\es{3.2}{-2}$              & $\es{7.1}{-3}$            & $\es{1.1}{-2}$             & $\es{1.4}{-2}$             & -                                 & -                      \\
Cas A$^{\ddagger\ddagger}$                & $\es{8.0}{-1}$              &   $0.00$       &    $\es{8.0}{-1}$         &  $0.00$            & -                                 & -                      \\
\hline
\enddata
\tablecomments{Comparisons with Observations and other simulation results. The out-of-cone, $\mathrm{E}_{out\-of\-cone}$, and in-cone, $\mathrm{E}_{out\-of\-cone}$, energies correspond to the explosion energies of the models used for these dust mass estimations of aSNe, HNe, and Cas A like comparisons using our database. We give the total dust mass, $\mathrm{M}_{total\ dust}$, carbonaceous dust mass, $\mathrm{M}_{\mathrm{C}(s)}$, forsterite dust mass, $\mathrm{M}_{\forsterite(s)}$, and alumina dust mass $\mathrm{M}_{\alumina(s)}$ to compare to some of the higher yield grains seen in both observations and simulations. It should be noted that we are using data evolved out to 1157 days.\\
$^{\dagger}$ Observed dust mass at 1153 days, \citet{Wesson2014,wesson2015timing}.\\
$^{\dagger\dagger}$ Observed, \citet{Priestley2019MNRAS.485..440P}.\\
$^{\ddagger}$ Simulated, 4000 days, \citet{Biscaro2016}.\\
$^{\ddagger\ddagger}$ \citet{delooze_casa_2019}. \\
$^{a}$ Parameter fit f=3,$\Theta$=40, \citet{Hungerford2005ApJ...635..487H}.\\
$^{b}$ Parameter fit f=2,$\Theta$=20, \citet{Hungerford2005ApJ...635..487H}.\\
$^{c}$ Parameter fit f=5,$\Theta$=20, \citet{Hungerford2005ApJ...635..487H}.\\
$^{d}$ $\Theta$=15.\\
$^{e}$ $\Theta$=10.\\
$^{f}$ $\Theta$=40.\\
}
\end{deluxetable}

%% file: Figures/dustvm.tex
\begin{figure}[H]
    \centering
    \includegraphics[scale=0.95,angle=0]{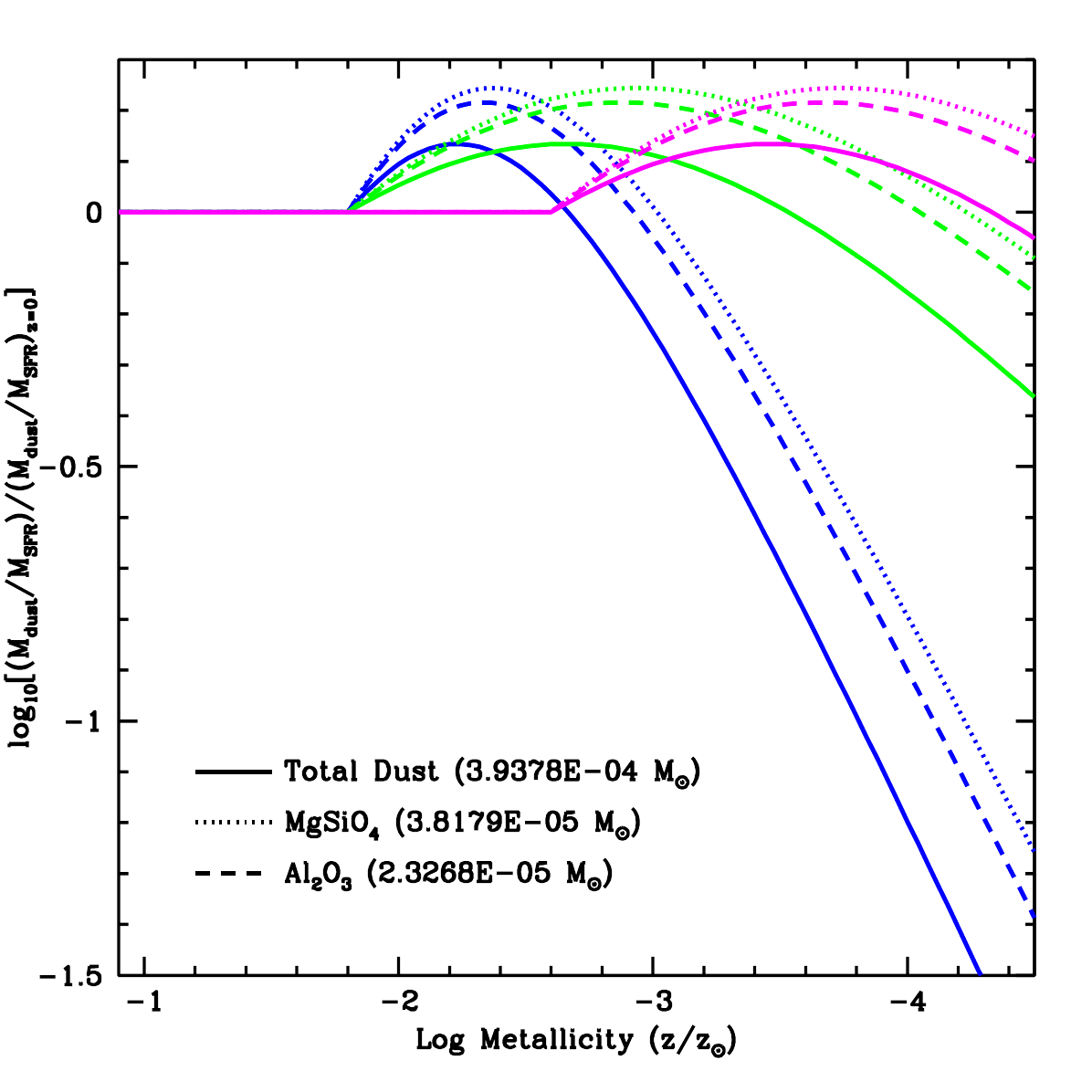}
    \caption{Dust production (relative to the redshift 0, solar metallicity dust production) versus metallicity for three different evolutionary models for the initial mass function.  In these models, we vary the onset of the flattening of the IMF from an early evolution  starting at a metallicity of 2.5\% solar (roughly a redshift of 5) and late evolution at 0.3\% solar metallicity (roughly a redshift of 7).  For our early evolution models, we include two scenarios where the initial mass function flattens at two different rates.  These models are designed to show a range of possible dust evolution scenarios.  The solid, dotted and dashed curves correspond to the total, MgSiO$_4$ and Al$_2$O$_3$ dust production.}
    \label{fig:dustvm}
\end{figure}

%% file: appendix.tex
\section{Nucleation}\label{appx:nucl}
 The modified steaty-state rate of this nucleation reaction Eq. (\ref{eq:reaction_diagram_nucleation}), is given by
 \begin{equation}\label{eq:steady_nucleation}
 J_s = \gamma \Omega_0 \sqrt{\frac{ 2 \sigma}{\pi m_1}} c_1^2 \Pi \exp \left(- \frac{4}{27} \frac{\mu^3}{\left(\ln S\right)^2} \right)
 \end{equation}
 where $\gamma$ is a sticking probability\footnote{We assume $\gamma = 1$}, $\Omega_0 = 4 \pi a_0 ^3 / 3$ is the volume per key-species of the cluster, $\sigma$ is the bulk-derived surface tension, $m_1$ is the mass of the attaching monomer, $S$ is the \textit{supersaturation ratio} (hereafter \textit{saturation}), $\mu = 4 \pi a_0^2 \sigma / kT$,  and $\Pi$ is a correction factor defined as
 \begin{equation}\label{eq:correct_nucleation}
 \Pi = \left[\frac{\prod_{k=1}^i (c^\fancylet{A}_k / c_1)^{\nu_k}}{\prod_{k=1}^j (c^\fancylet{B}_k / c_1)^{\eta_k}}\right]^{\frac{1}{\omega}}
 \end{equation}
 where $c^A, c^B$ are the concentrations of reactants, products respectively, $c_1$ is the key-species concentration in the vapor, and $w = 1 + \sum_{k=1}^i \nu_k - \sum_{k=1}^j \eta_k$. Values for parameters $a_0$, $\sigma$ of each key species is given in Table \ref{tab:Nozawa2003Table}.
 
 The thermodynamics of phase-change are determined by the saturation $S$, which is evaluated w.r.t. the key species as 
 \begin{equation}\label{eq:keyspec_S}
 \ln S = \ln \frac{p_1}{p_{1,s}} = -\frac{\Delta G}{kT} - \left\{\ln \frac{p_1}{p_s} - \ln \left[\frac{\prod_{k=1}^i (p_k^\fancylet{A} / p_s)^{\nu_k}}{\prod_{k=1}^j (p_k^\fancylet{B} / p_s)^{\eta_k}}\right] \right\}
 \end{equation}
 where $p_{1,s}$ is the vapor pressure. The thermodynamic potential $\Delta G$ of a reaction is determined using a two-parameter data fit $\Delta G/kT = -A/T + B$, where $A, B$ are derived using the table of NASA coefficients \citet{mcbride1993coefficients} for the component species. The values used here are given in Table \ref{tab:Nozawa2003Table}.  
 
 In a Langranian cell of volume $V(t)$ the concentration of grains composed of $n$ monomers of the key species is given by $c_{n}(t) = N_{n}(t) / V(t)$, where $N_{n}(t)$ is the total number of $n$-mers. $c_{1}(t)$ represents the vapor-phase concentration of key species monomers. For convenience, let us introduce the \textit{nominal} concentration $\cbar_{n}$ defined as the concentration of monomers that would result if no dust formation occurred. By definition, then

\begin{equation} \label{eq:nucl_undepleted}
    c_{n}(t_0) V(t_0) = \cbar_{n}(t) V(t)
\end{equation}
where $t_0$ is the initial value of time. Eq. (\ref{eq:nucl_undepleted}) simply states that, without any nucleation depletion, the number of $n$-mers is conserved. Mass conservation of the key species can then be written as

\begin{equation} \label{eq:nucl_massconserve}
    \cbar_{1} V - c_{1} V = \sum_{n=2}^{n_{*}-1} n c_{n} V + \int_{t_{0}}^{t} V(t')J_{n_{*}}(t')\frac{a^3(t,t')}{a_{0}^{3}} dt'
\end{equation}
where $n_{*}$ is the \textit{critical size}, $a(t,t')$ is the radius of the of the grain nucleated at $t'$ measured at $t$, and $a_0$ is the monomer radius. The summation on the RHS counts all current $n$-mers formed up to the critical cluster, and the integration accounts for the nucleation of growth of all $n$-mers since $t_0$.

Instead of following the detailed kinetics of pre-critical $n$-mers, we assume all grains form from the vapor as critical clusters and use the modified form of steady-state nucleation rate Eq. (\ref{eq:steady_nucleation}).
 Eq. (\ref{eq:nucl_massconserve}) simplifies to

 \begin{equation} \label{eq:nucl_massconserve_simple}
     \cbar_{1} V - c_{1} V = \int_{t_{0}}^{t} V(t') J_{*}(t')\frac{a^3(t,t')}{a_{0}^{3}} dt'
 \end{equation}
where $J_{*}(t)$ is the modified steady-state rate given by Eq. (\ref{eq:correct_nucleation}), where the $*$ subscript indicates that we are nucleating critical size clusters. Let $I_{*} = J_{*} / \cbar_1$, divide by $\cbar_{1}(t)V(t)$ and let

\begin{equation} \label{eq:nucl_k3}
    K_{3} = \int_{t_{0}}^{t} I_*(t')\frac{a^3(t,t')}{a_{0}^{3}} dt'
\end{equation} 
we arrive at the simple equation for mass conservation
\begin{equation} \label{eq:nucl_massconserve_simple_again}
    1 - \frac{c_{1}}{\cbar_{1}} = 1 - Y_{1} = K_{3}
\end{equation}
where $Y_{1} = c_{1}/\cbar_{1}$ is the normalized concentration of key species monomers.

The integral equation Eq. (\ref{eq:nucl_massconserve_simple_again}) is solved by a transformation into a set of first-order differential equations (ODEs). Repeated differentiation of Eq. (\ref{eq:nucl_k3}) leads to

\begin{equation} \label{eq:nucl_moments}
\begin{split}
\frac{d K_i }{d t} = \begin{cases}
    I_*(t)n_*^{\frac{i}{3}} + \frac{i}{a_0} \left( \frac{d a}{d t} \right) K_{i-1} & \text{for $i = 1 \dots 3$} \\
    I_*(t) & \text{for $i = 0$}
\end{cases}
\end{split}
\end{equation}
These equations are coupled to that of grain growth

\begin{equation} \label{eq:nucl_graingrow}
    \frac{da}{dt} = \gamma \Omega_0 \sqrt{ \frac{kT}{2 \pi m_1}} c_1 \left(1 - \frac{1}{S}\right)
\end{equation}
to allow the determination of $K_{3}$, and the grain concentration $Y_1$ immediately follows from Eq. (\ref{eq:nucl_massconserve_simple_again}).

The concentrations of non-key species due to nucleation are determined by the rate of key-species as

\begin{equation} \label{eq:nucl_nonks_evo}
\begin{split}
    Y_k^\fancylet{A} = \frac{c_k^\fancylet{A}}{\cbar_1} = \frac{\cbar_k^\fancylet{A}}{\cbar_1} - \nu_k^\fancylet{A} (1-Y_1)\\
    Y_k^\fancylet{B} = \frac{c_k^\fancylet{B}}{\cbar_1} = \frac{\cbar_k^\fancylet{B}}{\cbar_1} + \eta_k^\fancylet{B} (1 - Y_1)
\end{split}
\end{equation}
where $\fancylet{A}, \fancylet{B}$ identify reactant, product species as in the reaction given by Eq. (\ref{eq:reaction_diagram_nucleation}). Further, various grain properties naturally arise from inspection of the moments $K_i$:

\begin{align} \label{eq:nucl_moment_translate}
        N_{dust} &= \cbar_{1} K_0 \\
        c_{dust} &= \cbar_{1} K_3 \\
        \hat{r}_{dust} &= a_0 \left( K_3/K_0 \right)^{1/3}
        \end{align}
where $N_{dust}$ is the total number of grains, $c_{dust}$ is the concentration of grains, and $\hat{r}_{dust}$ is the average grain radius.

\section{Data tables}
\input{Tables/fryer_model_table}

\input{Tables/isotopes}
\clearpage
\input{Tables/grain_reactions_table}
\clearpage
\input{Tables/dust_mass_table}

%% file: Tables/fryer_model_table.tex
\startlongtable
\begin{deluxetable}{c|ccccccc}
\tablecaption{List of models used from \cite{fryer2018}.
\label{tab:fryer-models}}
\tablehead{\colhead{Model} & \colhead{$\Mprog$} & \colhead{$\Mbounce$} & \colhead{$\Minj$} & \colhead{$\tinj$} & \colhead{$\Einj$} & \colhead{$\Eexp$} & \colhead{$\Mrem$}\\
 & $(\Msol)$ & $(\Msol)$ & $(\Msol)$ & $(\mathrm{s})$ & $\tens{51}\ \uerg$ & $\tens{51}\ \uerg$ & $(\Msol)$ 
}
\colnumbers
\startdata
$\mathrm{M15aE0.34}$  &  $15$    &   $1.30$    &   $0.3 $ &   $0.1$  &   $3 $   &   $0.34$ &   $1.94$ \\  
$\mathrm{M15aE0.54}$  &  $15$    &   $1.30$    &   $0.3 $ &   $0.1$  &   $4 $   &   $0.54$ &   $1.91$ \\  
$\mathrm{M15aE0.82}$  &  $15$    &   $1.30$    &   $0.3 $ &   $0.1$  &   $5 $   &   $0.82$ &   $1.88$ \\  
$\mathrm{M15aE2.47}$  &  $15$    &   $1.30$    &   $0.3 $ &   $0.1$  &   $9 $   &   $2.47$ &   $1.52$ \\  
$\mathrm{M15aE4.79}$  &  $15$    &   $1.30$    &   $0.3 $ &   $0.4$  &   $20$   &   $4.79$ &   $1.50$ \\  
$\mathrm{M15bE0.30}$  &  $15$    &   $1.30$    &   $0.02$ &   $0.4$  &   $3 $   &   $0.3 $ &   $1.71$ \\  
$\mathrm{M15bE0.52}$  &  $15$    &   $1.30$    &   $0.02$ &   $0.2$  &   $5 $   &   $0.52$ &   $1.71$ \\  
$\mathrm{M15bE0.74}$  &  $15$    &   $1.30$    &   $0.02$ &   $0.4$  &   $3 $   &   $0.74$ &   $1.73$ \\  
$\mathrm{M15bE0.82}$  &  $15$    &   $1.30$    &   $0.02$ &   $0.2$  &   $6 $   &   $0.82$ &   $1.71$ \\  
$\mathrm{M15bE0.89}$  &  $15$    &   $1.30$    &   $0.02$ &   $0.2$  &   $5 $   &   $0.89$ &   $1.74$ \\  
$\mathrm{M15bE0.92}$  &  $15$    &   $1.30$    &   $0.02$ &   $0.3$  &   $4 $   &   $0.92$ &   $1.75$ \\  
$\mathrm{M15bE1.69}$  &  $15$    &   $1.30$    &   $0.02$ &   $0.2$  &   $10$   &   $1.69$ &   $1.52$ \\  
$\mathrm{M15bE2.63}$  &  $15$    &   $1.30$    &   $0.02$ &   $0.2$  &   $20$   &   $2.63$ &   $1.53$ \\  
$\mathrm{M15bE10.7}$  &  $15$    &   $1.30$    &   $0.02$ &   $0.2$  &   $80$   &   $10.7$ &   $1.53$ \\  
$\mathrm{M15cE2.06}$  &  $15$    &   $1.30$    &   $0.1 $ &   $0.3$  &   $15$   &   $2.06$ &   $1.59$ \\  
$\mathrm{M15cE1.94}$  &  $15$    &   $1.30$    &   $0.1 $ &   $0.3$  &   $12$   &   $1.94$ &   $1.61$ \\  
$\mathrm{M15cE1.90}$  &  $15$    &   $1.30$    &   $0.1 $ &   $0.3$  &   $10$   &   $1.90$ &   $1.62$ \\  
$\mathrm{M15cE1.86}$  &  $15$    &   $1.30$    &   $0.1 $ &   $0.3$  &   $9 $   &   $1.86$ &   $1.63$ \\  
$\mathrm{M15cE2.24}$  &  $15$    &   $1.30$    &   $0.1 $ &   $0.3$  &   $25$   &   $2.24$ &   $1.56$ \\  
$\mathrm{M15cE2.60}$  &  $15$    &   $1.30$    &   $0.1 $ &   $0.3$  &   $45$   &   $2.60$ &   $1.52$ \\  
$\mathrm{M15cE3.43}$  &  $15$    &   $1.30$    &   $0.1 $ &   $0.3$  &   $90$   &   $3.43$ &   $1.51$ \\  
\hline
$\mathrm{M20aE0.53}$ &  $20$   &   $1.56$    &   $0.1$   &   $0.50$  &   $4  $   &   $0.53$   &   $3.40$ \\ 
$\mathrm{M20aE0.65}$ &  $20$   &   $1.56$    &   $0.1$   &   $0.12$  &   $4  $   &   $0.65$   &   $3.03$ \\ 
$\mathrm{M20aE0.81}$ &  $20$   &   $1.56$    &   $0.1$   &   $0.12$  &   $7  $   &   $0.81$   &   $2.70$ \\ 
$\mathrm{M20aE0.85}$ &  $20$   &   $1.56$    &   $0.1$   &   $0.50$  &   $7  $   &   $0.85$   &   $2.62$ \\ 
$\mathrm{M20aE1.39}$ &  $20$   &   $1.56$    &   $0.1$   &   $0.12$  &   $10 $   &   $1.39$   &   $1.93$ \\ 
$\mathrm{M20aE1.47}$ &  $20$   &   $1.56$    &   $0.1$   &   $0.50$  &   $10 $   &   $1.47$   &   $2.23$ \\ 
$\mathrm{M20aE2.43}$ &  $20$   &   $1.56$    &   $0.1$   &   $0.12$  &   $20 $   &   $2.43$   &   $1.86$ \\ 
$\mathrm{M20aE2.50}$ &  $20$   &   $1.56$    &   $0.1$   &   $0.50$  &   $20 $   &   $2.50$   &   $1.93$ \\ 
$\mathrm{M20aE4.15}$ &  $20$   &   $1.56$    &   $0.1$   &   $0.12$  &   $50 $   &   $4.15$   &   $1.85$ \\ 
$\mathrm{M20bE0.78}$ &  $20$   &   $1.56$    &   $0.2$   &   $0.12$  &   $5  $   &   $0.78$   &   $2.85$ \\ 
$\mathrm{M20bE1.04}$ &  $20$   &   $1.56$    &   $0.2$   &   $0.12$  &   $6  $   &   $1.04$   &   $2.47$ \\ 
$\mathrm{M20bE1.19}$ &  $20$   &   $1.56$    &   $0.2$   &   $0.12$  &   $8  $   &   $1.19$   &   $2.28$ \\ 
$\mathrm{M20bE1.52}$ &  $20$   &   $1.56$    &   $0.2$   &   $0.12$  &   $10 $   &   $1.52$   &   $1.97$ \\ 
$\mathrm{M20bE2.60}$ &  $20$   &   $1.56$    &   $0.2$   &   $0.12$  &   $25 $   &   $2.60$   &   $1.90$ \\ 
$\mathrm{M20bE4.33}$ &  $20$   &   $1.56$    &   $0.2$   &   $0.12$  &   $50 $   &   $4.33$   &   $1.87$ \\ 
$\mathrm{M20cE0.75}$ &  $20$   &   $1.47$    &   $0.1$   &   $0.5 $  &   $6  $   &   $0.75$   &   $2.76$ \\ 
$\mathrm{M20cE0.84}$ &  $20$   &   $1.47$    &   $0.1$   &   $0.5 $  &   $7  $   &   $0.84$   &   $2.62$ \\ 
$\mathrm{M20cE1.00}$ &  $20$   &   $1.47$    &   $0.1$   &   $0.5 $  &   $8  $   &   $1.00$   &   $2.35$ \\ 
$\mathrm{M20cE1.65}$ &  $20$   &   $1.47$    &   $0.1$   &   $0.5 $  &   $10 $   &   $1.65$   &   $1.78$ \\ 
$\mathrm{M20cE2.76}$ &  $20$   &   $1.47$    &   $0.1$   &   $0.5 $  &   $15 $   &   $2.76$   &   $1.76$ \\ 
$\mathrm{M20cE2.85}$ &  $20$   &   $1.47$    &   $0.1$   &   $0.5 $  &   $20 $   &   $2.85$   &   $1.74$ \\ 
$\mathrm{M20cE5.03}$ &  $20$   &   $1.47$    &   $0.1$   &   $0.5 $  &   $50 $   &   $5.03$   &   $1.74$ \\ 
$\mathrm{M20cE8.86}$ &  $20$   &   $1.47$    &   $0.1$   &   $0.5 $  &   $100$   &   $8.86$   &   $1.74$ \\ 
$\mathrm{M20dE4.33}$ &  $20$   &   $1.56$    &   $0.2$   &   $0.12$  &   $50 $   &   $4.33$   &   $1.87$ \\ 
$\mathrm{M20dE5.90}$ &  $20$   &   $1.47$    &   $0.2$   &   $0.5 $  &   $20 $   &   $5.9 $   &   $1.74$ \\ 
$\mathrm{M20dE18.1}$ &  $20$   &   $1.47$    &   $0.2$   &   $0.5 $  &   $50 $   &   $18.1$   &   $1.74$ \\ 
$\mathrm{M20dE64.5}$ &  $20$   &   $1.47$    &   $0.2$   &   $0.5 $  &   $75 $   &   $64.5$   &   $1.74$ \\ 
$\mathrm{M20dE78.9}$ &  $20$   &   $1.47$    &   $0.2$   &   $0.5 $  &   $100$   &   $78.9$   &   $1.74$ \\ 
$\mathrm{M20dE88.4}$ &  $20$   &   $1.47$    &   $0.2$   &   $0.5 $  &   $125$   &   $88.4$   &   $1.74$ \\ 
$\mathrm{M20dE124}$  &  $20$   &   $1.47$    &   $0.2$   &   $0.5 $  &   $150$   &   $124 $   &   $1.74$ \\ 
\hline
$\mathrm{M25aE0.99 }$  &  $25$   &   $1.83$   &   $0.1 $   &   $0.1 $  &  $5.0 $  &  $0.99$   &  $4.89$ \\ 
$\mathrm{M25aE1.57 }$  &  $25$   &   $1.83$   &   $0.1 $   &   $0.1 $  &  $10  $  &  $1.57$   &  $3.73$ \\ 
$\mathrm{M25aE4.73 }$  &  $25$   &   $1.83$   &   $0.1 $   &   $0.1 $  &  $20  $  &  $4.73$   &  $2.38$ \\ 
$\mathrm{M25aE6.17 }$  &  $25$   &   $1.83$   &   $0.1 $   &   $0.1 $  &  $35  $  &  $6.17$   &  $2.38$ \\ 
$\mathrm{M25aE7.42 }$  &  $25$   &   $1.83$   &   $0.1 $   &   $0.1 $  &  $50  $  &  $7.42$   &  $2.37$ \\ 
$\mathrm{M25aE14.8 }$  &  $25$   &   $1.83$   &   $0.1 $   &   $0.1 $  &  $100 $  &  $14.8$   &  $2.35$ \\ 
$\mathrm{M25bE8.40 }$  &  $25$   &   $1.83$   &   $0.02$   &   $0.28$  &  $50.0$  &  $8.40$   &  $2.38$ \\ 
$\mathrm{M25bE9.73 }$  &  $25$   &   $1.83$   &   $0.02$   &   $0.69$  &  $100 $  &  $9.73$   &  $2.35$ \\ 
$\mathrm{M25bE18.4 }$  &  $25$   &   $1.83$   &   $0.02$   &   $0.69$  &  $200 $  &  $18.4$   &  $2.35$ \\ 
$\mathrm{M25d3E0.89}$  &  $25$   &   $1.83$   &   $0.02$   &   $0.7 $  &  $7   $  &  $0.89$   &  $4.66$ \\ 
$\mathrm{M25d3E0.92}$  &  $25$   &   $1.83$   &   $0.02$   &   $0.7 $  &  $8   $  &  $0.92$   &  $1.84$ \\ 
$\mathrm{M25d3E1.04}$  &  $25$   &   $1.83$   &   $0.02$   &   $0.7 $  &  $10  $  &  $1.04$   &  $1.84$ \\ 
$\mathrm{M25d3E1.20}$  &  $25$   &   $1.83$   &   $0.02$   &   $0.7 $  &  $50  $  &  $1.20$   &  $1.84$ \\ 
$\mathrm{M25d2E2.53}$  &  $25$   &   $1.83$   &   $0.02$   &   $0.7 $  &  $20  $  &  $2.53$   &  $2.35$ \\ 
$\mathrm{M25d2E2.64}$  &  $25$   &   $1.83$   &   $0.02$   &   $0.7 $  &  $35  $  &  $2.64$   &  $2.35$ \\ 
$\mathrm{M25d2E2.78}$  &  $25$   &   $1.83$   &   $0.02$   &   $0.7 $  &  $50  $  &  $2.78$   &  $2.35$ \\ 
$\mathrm{M25d2E3.07}$  &  $25$   &   $1.83$   &   $0.02$   &   $0.7 $  &  $100 $  &  $3.07$   &  $1.83$ \\ 
$\mathrm{M25d1E3.30}$  &  $25$   &   $1.83$   &   $0.02$   &   $0.7 $  &  $25  $  &  $3.30$   &  $2.35$ \\ 
$\mathrm{M25d1E4.72}$  &  $25$   &   $1.83$   &   $0.02$   &   $0.7 $  &  $50  $  &  $4.72$   &  $2.35$ \\ 
$\mathrm{M25d1E7.08}$  &  $25$   &   $1.83$   &   $0.02$   &   $0.7 $  &  $100 $  &  $7.08$   &  $2.35$ \\ 
\enddata
\tablecomments{List of supernova models used organized by progenitor mass denoted by uppercase "M" with progenitor mass in the model name. For each progenitor mass, lowercase alphabetic characters denote supernova engine subgroups. Subgroups are order by increasing explosion energy denoted with uppercase "E" and the explosion energy in the model name. Table columns: 1) Model, 2) Progenitor mass, $\Mprog$, 3) Shock rebound mass, $\Mbounce$, 4) Injection mass, $\Minj$, 5) Injection time, $\tinj$, 6) Injection energy, $\Einj$, 7) Final explosion energy, $\Eexp$, 8) Mass of remnant, $\Mrem$.}
\end{deluxetable}

%% file: Tables/isotopes.tex
\startlongtable
\begin{deluxetable}{|c|c|c|c|c|c|c|c|c|c|}
\tablecaption{Table of Isotopes \label{tab:isoTable}}
\tablehead{}
\startdata
H   2 & O  21 & MG 43 & P  37 & CL 43 & K  40 & CA 72 & TI 60 & CR 42 & MN 68 \\ 
HE  3 & O  22 & MG 44 & P  38 & CL 44 & K  41 & CA 73 & TI 61 & CR 43 & MN 69 \\ 
HE  4 & F  20 & MG 45 & P  39 & CL 45 & K  42 & SC 32 & TI 62 & CR 44 & MN 70 \\ 
BE  7 & F  21 & MG 46 & P  40 & CL 46 & K  43 & SC 33 & TI 63 & CR 45 & MN 71 \\ 
B   8 & F  22 & MG 47 & P  41 & CL 47 & K  44 & SC 34 & TI 64 & CR 46 & MN 72 \\ 
LI  7 & F  23 & AL 21 & P  42 & CL 48 & K  45 & SC 35 & TI 65 & CR 47 & MN 73 \\ 
C  11 & F  24 & AL 22 & P  43 & CL 49 & K  46 & SC 36 & TI 66 & CR 48 & MN 74 \\ 
B  11 & F  25 & AL 23 & P  44 & CL 50 & K  47 & SC 37 & TI 67 & CR 49 & MN 75 \\ 
C  12 & F  26 & AL 24 & P  45 & CL 51 & K  48 & SC 38 & TI 68 & CR 50 & MN 76 \\ 
C  13 & NE 17 & AL 28 & P  46 & CL 52 & K  49 & SC 39 & TI 69 & CR 51 & MN 77 \\ 
N  13 & NE 18 & AL 29 & P  47 & CL 53 & K  50 & SC 40 & TI 70 & CR 52 & MN 78 \\ 
N  14 & NE 19 & AL 30 & P  48 & CL 54 & K  51 & SC 41 & TI 71 & CR 53 & MN 79 \\ 
C  14 & NE 23 & AL 31 & P  49 & CL 55 & K  52 & SC 42 & TI 72 & CR 54 & MN 80 \\ 
N  15 & NE 24 & AL 32 & P  50 & CL 56 & K  53 & SC 43 & TI 73 & CR 55 & MN 81 \\ 
O  16 & NE 25 & AL 33 & P  51 & CL 57 & K  54 & SC 44 & TI 74 & CR 56 & MN 82 \\ 
O  17 & NE 26 & AL 34 & P  52 & CL 58 & K  55 & SC 45 & TI 75 & CR 57 & MN 83 \\ 
O  18 & NE 27 & AL 35 & P  53 & CL 59 & K  56 & SC 46 & TI 76 & CR 58 & MN 84 \\ 
F  17 & NE 28 & AL 36 & P  54 & CL 60 & K  57 & SC 47 & TI 77 & CR 59 & MN 85 \\ 
F  18 & NE 29 & AL 37 & P  55 & CL 61 & K  58 & SC 48 & TI 78 & CR 60 & MN 86 \\ 
F  19 & NE 30 & AL 38 & P  56 & CL 62 & K  59 & SC 49 & TI 79 & CR 61 & MN 87 \\ 
NE 20 & NE 31 & AL 39 & P  57 & CL 63 & K  60 & SC 50 & TI 80 & CR 62 & MN 88 \\ 
NE 21 & NE 32 & AL 40 & S  25 & AR 27 & K  61 & SC 51 & V  36 & CR 63 & MN 89 \\ 
NE 22 & NE 33 & AL 41 & S  26 & AR 28 & K  62 & SC 52 & V  37 & CR 64 & FE 42 \\ 
NA 22 & NE 34 & AL 42 & S  27 & AR 29 & K  63 & SC 53 & V  38 & CR 65 & FE 43 \\ 
NA 23 & NE 35 & AL 43 & S  28 & AR 30 & K  64 & SC 54 & V  39 & CR 66 & FE 44 \\ 
MG 23 & NE 36 & AL 44 & S  29 & AR 31 & K  65 & SC 55 & V  40 & CR 67 & FE 45 \\ 
MG 24 & NE 37 & AL 45 & S  30 & AR 32 & K  66 & SC 56 & V  41 & CR 68 & FE 46 \\ 
MG 25 & NE 38 & AL 46 & S  32 & AR 33 & K  67 & SC 57 & V  42 & CR 69 & FE 47 \\ 
MG 26 & NE 39 & AL 47 & S  33 & AR 34 & K  68 & SC 58 & V  43 & CR 70 & FE 48 \\ 
AL 26 & NE 40 & AL 48 & S  34 & AR 35 & K  69 & SC 59 & V  44 & CR 71 & FE 49 \\ 
AL 27 & NE 41 & AL 49 & S  35 & AR 36 & K  70 & SC 60 & V  45 & CR 72 & FE 50 \\ 
SI 27 & NA 19 & AL 50 & S  36 & AR 37 & CA 30 & SC 61 & V  46 & CR 73 & FE 51 \\ 
SI 28 & NA 20 & AL 51 & S  37 & AR 38 & CA 31 & SC 62 & V  47 & CR 74 & FE 52 \\ 
SI 29 & NA 24 & SI 22 & S  38 & AR 39 & CA 32 & SC 63 & V  48 & CR 75 & FE 53 \\ 
SI 30 & NA 25 & SI 23 & S  39 & AR 40 & CA 33 & SC 64 & V  49 & CR 76 & FE 54 \\ 
P  31 & NA 26 & SI 24 & S  40 & AR 41 & CA 34 & SC 65 & V  50 & CR 77 & FE 55 \\ 
S  31 & NA 27 & SI 25 & S  41 & AR 42 & CA 35 & SC 66 & V  51 & CR 78 & FE 56 \\ 
BE  8 & NA 28 & SI 26 & S  42 & AR 43 & CA 36 & SC 67 & V  52 & CR 79 & FE 57 \\ 
O  14 & NA 29 & SI 31 & S  43 & AR 44 & CA 37 & SC 68 & V  53 & CR 80 & FE 58 \\ 
O  15 & NA 30 & SI 32 & S  44 & AR 45 & CA 38 & SC 69 & V  54 & CR 81 & FE 59 \\ 
NA 21 & NA 31 & SI 33 & S  45 & AR 46 & CA 39 & SC 70 & V  55 & CR 82 & FE 60 \\ 
AL 25 & NA 32 & SI 34 & S  46 & AR 47 & CA 40 & SC 71 & V  56 & CR 83 & FE 61 \\ 
P  29 & NA 33 & SI 35 & S  47 & AR 48 & CA 41 & SC 72 & V  57 & CR 84 & FE 62 \\ 
P  30 & NA 34 & SI 36 & S  48 & AR 49 & CA 42 & SC 73 & V  58 & CR 85 & FE 63 \\ 
PB 206 & NA 35 & SI 37 & S  49 & AR 50 & CA 43 & SC 74 & V  59 & CR 86 & FE 64 \\ 
PB 207 & NA 36 & SI 38 & S  50 & AR 51 & CA 44 & SC 75 & V  60 & MN 40 & FE 65 \\ 
BI 211 & NA 37 & SI 39 & S  51 & AR 52 & CA 45 & SC 76 & V  61 & MN 41 & FE 66 \\ 
PO 210 & NA 38 & SI 40 & S  52 & AR 53 & CA 46 & TI 34 & V  62 & MN 42 & FE 67 \\ 
H   3 & NA 39 & SI 41 & S  53 & AR 54 & CA 47 & TI 35 & V  63 & MN 43 & FE 68 \\ 
HE  6 & NA 40 & SI 42 & S  54 & AR 55 & CA 48 & TI 36 & V  64 & MN 44 & FE 69 \\ 
LI  8 & NA 41 & SI 43 & S  55 & AR 56 & CA 49 & TI 37 & V  65 & MN 45 & FE 70 \\ 
LI  9 & NA 42 & SI 44 & S  56 & AR 57 & CA 50 & TI 38 & V  66 & MN 46 & FE 71 \\ 
BE 10 & NA 43 & SI 45 & S  57 & AR 58 & CA 51 & TI 39 & V  67 & MN 47 & FE 72 \\ 
BE 11 & NA 44 & SI 46 & S  58 & AR 59 & CA 52 & TI 40 & V  68 & MN 48 & FE 73 \\ 
BE 12 & MG 20 & SI 47 & S  59 & AR 60 & CA 53 & TI 41 & V  69 & MN 49 & FE 74 \\ 
B  12 & MG 21 & SI 48 & S  60 & AR 61 & CA 54 & TI 42 & V  70 & MN 50 & FE 75 \\ 
B  13 & MG 22 & SI 49 & CL 26 & AR 62 & CA 55 & TI 43 & V  71 & MN 51 & FE 76 \\ 
B  14 & MG 27 & SI 50 & CL 27 & AR 63 & CA 56 & TI 44 & V  72 & MN 52 & FE 77 \\ 
C  15 & MG 28 & SI 51 & CL 28 & AR 64 & CA 57 & TI 45 & V  73 & MN 53 & FE 78 \\ 
C  16 & MG 29 & SI 52 & CL 29 & AR 65 & CA 58 & TI 46 & V  74 & MN 54 & FE 79 \\ 
C  17 & MG 30 & SI 53 & CL 30 & AR 66 & CA 59 & TI 47 & V  75 & MN 55 & FE 80 \\ 
C  18 & MG 31 & SI 54 & CL 31 & AR 67 & CA 60 & TI 48 & V  76 & MN 56 & FE 81 \\ 
N  11 & MG 32 & P  23 & CL 32 & K  29 & CA 61 & TI 49 & V  77 & MN 57 & FE 82 \\ 
N  12 & MG 33 & P  24 & CL 33 & K  30 & CA 62 & TI 50 & V  78 & MN 58 & FE 83 \\ 
N  16 & MG 34 & P  25 & CL 34 & K  31 & CA 63 & TI 51 & V  79 & MN 59 & FE 84 \\ 
N  17 & MG 35 & P  26 & CL 35 & K  32 & CA 64 & TI 52 & V  80 & MN 60 & FE 85 \\ 
N  18 & MG 36 & P  27 & CL 36 & K  33 & CA 65 & TI 53 & V  81 & MN 61 & FE 86 \\ 
N  19 & MG 37 & P  28 & CL 37 & K  34 & CA 66 & TI 54 & V  82 & MN 62 & FE 87 \\ 
N  20 & MG 38 & P  32 & CL 38 & K  35 & CA 67 & TI 55 & V  83 & MN 63 & FE 88 \\ 
N  21 & MG 39 & P  33 & CL 39 & K  36 & CA 68 & TI 56 & CR 38 & MN 64 & FE 89 \\ 
O  13 & MG 40 & P  34 & CL 40 & K  37 & CA 69 & TI 57 & CR 39 & MN 65 & FE 90 \\ 
O  19 & MG 41 & P  35 & CL 41 & K  38 & CA 70 & TI 58 & CR 40 & MN 66 & FE 91 \\ 
O  20 & MG 42 & P  36 & CL 42 & K  39 & CA 71 & TI 59 & CR 41 & MN 67 & FE 92 \\ 
\enddata
\tablecomments{List of Isotopes used in the abundances of each model}
\end{deluxetable}

%% file: Tables/grain_reactions_table.tex
\startlongtable
\begin{deluxetable}{c|cccccc}
\tablecaption{Dust grain reaction table taken from \cite{Nozawa_2003}. \label{tab:Nozawa2003Table}}
\tablehead{
\colhead{Grain} & \colhead{Key Species} & \colhead{Formula} & \colhead{$\mathrm{A} / \tens{4} (\uT)$} & \colhead{B} & \colhead{$\sigma (\uerg/ \angstrom^2) $} & \colhead{$a_{0} (\angstrom)$}\\
}
\colnumbers
\startdata
$\carb$&$\mathrm{C(g)}$&$\mathrm{C(g)}\rightarrow\carb$&$8.64726$&$19.0422$&$1400.0$&$1.281$\\
$\SiC$&$\mathrm{Si(g),C(g)}$&$\mathrm{Si(g)+C(g)}\rightarrow\SiC$&$14.8934$&$37.3825$&$1800.0$&$1.702$\\
$\TiC$&$\mathrm{Ti(g),C(g)}$&$\mathrm{Ti(g)+C(g)}\rightarrow\TiC$&$16.4696$&$37.2301$&$1242.0$&$1.689$\\
$\sili$&$\mathrm{Si(g)}$&$\mathrm{Si(g)}\rightarrow\sili$&$5.36975$&$17.4349$&$800.0$&$1.684$\\
$\enstatite\mathrm{(s)}$&$\mathrm{Mg(g),SiO(g)}$&$\mathrm{Mg(g)+SiO(g)+2O(g)}\rightarrow\enstatite\mathrm{(s)}$&$25.0129$&$72.0015$&$400.0$&$2.319$\\
$\forsterite\mathrm{(s)}$&$\mathrm{Mg(g)}$&$\mathrm{2Mg(g)+SiO(g)+3O(g)}\rightarrow\forsterite\mathrm{(s)}$&$18.6200$&$52.4336$&$436.0$&$2.055$\\
$\forsterite\mathrm{(s)}$&$\mathrm{SiO(g)}$&$\mathrm{2Mg(g)+SiO(g)+3O(g)}\rightarrow\forsterite\mathrm{(s)}$&$37.2400$&$104.872$&$436.0$&$2.589$\\
$\silica\mathrm{(s)}$&$\mathrm{SiO\mathrm{(g)}}$&$\mathrm{SiO\mathrm{(g)}+O(g)}\rightarrow\silica\mathrm{(s)}$&$12.6028$&$38.1507$&$605.0$&$2.080$\\
$\alumina\mathrm{(s)}$&$\mathrm{Al(g)}$&$\mathrm{2Al(g)+3O(g)}\rightarrow\alumina\mathrm{(s)}$&$18.4788$&$45.3543$&$690.0$&$1.718$\\
$\magnesia\mathrm{(s)}$&$\mathrm{Mg(g)}$&$\mathrm{Mg(g)+O(g)}\rightarrow\magnesia\mathrm{(s)}$&$11.9237$&$33.1593$&$1100.0$&$1.646$\\
$\mathrm{FeO\mathrm{(s)}}$&$\mathrm{Fe(g)}$&$\mathrm{Fe(g)+O(g)}\rightarrow\mathrm{FeO\mathrm{(s)}}$&$11.1290$&$31.9850$&$580.0$&$1.682$\\
$\magnetite\mathrm{(s)}$&$\mathrm{Fe(g)}$&$\mathrm{3Fe(g)+4O(g)}\rightarrow\magnetite\mathrm{(s)}$&$13.2889$&$39.1687$&$400.0$&$1.805$\\
$\mathrm{FeS\mathrm{(s)}}$&$\mathrm{Fe(g),S(g)}$&$\mathrm{Fe(g)+S(g)}\rightarrow\mathrm{FeS\mathrm{(s)}}$&$9.31326$&$30.7771$&$380.0$&$1.932$\\
$\mathrm{Ti\mathrm{(s)}}$&$\mathrm{Ti(g)}$&$\mathrm{Ti(g)}\rightarrow\mathrm{Ti\mathrm{(s)}}$&$5.58902$&$16.6071$&$1510.0$&$1.615$\\
$\mathrm{V\mathrm{(s)}}$&$\mathrm{V(g)}$&$\mathrm{V(g)}\rightarrow\mathrm{V\mathrm{(s)}}$&$6.15394$&$17.8702$&$1697.0$&$1.490$\\
$\mathrm{Cr\mathrm{(s)}}$&$\mathrm{Cr(g)}$&$\mathrm{Cr(g)}\rightarrow\mathrm{Cr\mathrm{(s)}}$&$4.67733$&$16.7596$&$1880.0$&$1.421$\\
$\mathrm{Co\mathrm{(s)}}$&$\mathrm{Co(g)}$&$\mathrm{Co(g)}\rightarrow\mathrm{Co\mathrm{(s)}}$&$5.03880$&$16.8372$&$1936.0$&$1.383$\\
$\mathrm{Fe\mathrm{(s)}}$&$\mathrm{Fe(g)}$&$\mathrm{Fe(g)}\rightarrow\mathrm{Fe\mathrm{(s)}}$&$4.84180$&$16.5566$&$1800.0$&$1.411$\\
$\mathrm{Ni\mathrm{(s)}}$&$\mathrm{Ni(g)}$&$\mathrm{Ni(g)}\rightarrow\mathrm{Ni\mathrm{(s)}}$&$5.09310$&$17.1559$&$1924.0$&$1.377$\\
$\mathrm{Cu\mathrm{(s)}}$&$\mathrm{Cu(g)}$&$\mathrm{Cu(g)}\rightarrow\mathrm{Cu\mathrm{(s)}}$&$3.97955$&$14.9083$&$1300.0$&$1.412$\\
\enddata
\tablecomments{List of grain reactions. The parameters $A, B$ are parameters for finding free energy\tablenotemark{a} $-\Delta G / k T = -A / T + B$\ $\sigma$ gives experimentally determined surface tensions, and $a0$ is the expected monomer size.}
\tablenotetext{a}{These values can be found using the NASA coefficients for individual species \cite{mcbride1993coefficients}. In brief, $\Delta G = \Delta H - T \Delta S$. The polynomials given in \cite{mcbride1993coefficients} provide $\Delta H, \Delta S$ for the species to determine $\Delta G$. Values presented here are fit to a simpler form $-A / T + B$, although this is not required.}
\end{deluxetable}

%% file: Tables/dust_mass_table.tex


\startlongtable
\begin{deluxetable}{c|cccccccc}
\tablecaption{Dust mass for specific species and total dust mass produced per model by 1157 days after explosion.
\label{tab:dust-masses}}
\tablehead{
 \colhead{Dust Species} \vline & \colhead{Models}
}
\startdata
                 & M15aE0.34        & M15aE0.54        & M15aE0.82        & M15aE2.47        & M15aE4.79        & M15bE0.3          & M15bE0.52        & M15bE0.74        \\  
\hline
$\mathrm{C}$     & $\es{1.19}{-15}$ & $\es{6.96}{-7}$  & $\es{2.26}{-2}$  & $\es{4.47}{-2}$  & $\es{4.64}{-2}$  & $\es{1.27}{-5}$   & $\es{6.98}{-4}$  & $\es{2.14}{-2}$  \\ 
$\mathrm{SiC}$   & $\es{6.98}{-34}$ & $\es{6.17}{-15}$ & $\es{1.11}{-12}$ & $\es{2.67}{-12}$ & $\es{2.26}{-10}$ & $\es{7.34}{-14}$  & $\es{2.20}{-12}$ & $\es{3.45}{-12}$ \\ 
$\mathrm{TiC}$   &  0               & $\es{4.01}{-30}$ & $\es{1.47}{-12}$ & $\es{3.83}{-12}$ & $\es{7.53}{-13}$ &  0                & $\es{5.26}{-15}$ & $\es{9.95}{-13}$ \\ 
$\mathrm{Si}$    &  0               &  0               &  0               & $\es{2.77}{-10}$ & $\es{2.26}{-8}$  &  0                &  0               &  0               \\
$\silica$        & $\es{5.86}{-7}$  & $\es{3.72}{-7}$  & $\es{2.28}{-7}$  & $\es{8.15}{-8}$  & $\es{1.47}{-2}$  & $\es{6.99}{-7}$   & $\es{4.18}{-7}$  & $\es{2.69}{-7}$  \\
$\enstatite$     & $\es{3.43}{-6}$  & $\es{2.14}{-6}$  & $\es{1.28}{-6}$  & $\es{4.69}{-7}$  & $\es{9.30}{-4}$  & $\es{4.06}{-6}$   & $\es{2.38}{-6}$  & $\es{1.51}{-6}$  \\
$\forsterite$    & $\es{2.70}{-5}$  & $\es{1.68}{-5}$  & $\es{1.00}{-5}$  & $\es{3.71}{-6}$  & $\es{1.23}{-1}$  & $\es{3.21}{-5}$   & $\es{1.87}{-5}$  & $\es{1.19}{-5}$  \\
$\magnesia$      & $\es{6.42}{-8}$  & $\es{4.08}{-8}$  & $\es{2.51}{-8}$  & $\es{8.78}{-9}$  & $\es{3.46}{-8}$  & $\es{7.67}{-8}$   & $\es{4.61}{-8}$  & $\es{2.95}{-8}$  \\
$\alumina$       & $\es{1.26}{-6}$  & $\es{7.80}{-7}$  & $\es{4.77}{-7}$  & $\es{1.70}{-2}$  & $\es{1.71}{-2}$  & $\es{1.47}{-6}$   & $\es{8.57}{-7}$  & $\es{5.59}{-7}$  \\
$\mathrm{FeO}$   & $\es{7.39}{-6}$  & $\es{5.01}{-6}$  & $\es{3.25}{-6}$  & $\es{1.23}{-6}$  & $\es{4.66}{-7}$  & $\es{8.64}{-6}$   & $\es{5.58}{-6}$  & $\es{3.76}{-6}$  \\
$\magnetite$     & $\es{1.05}{-4}$  & $\es{7.07}{-5}$  & $\es{4.52}{-5}$  & $\es{1.73}{-5}$  & $\es{7.01}{-6}$  & $\es{1.22}{-4}$   & $\es{7.84}{-5}$  & $\es{5.27}{-5}$  \\
$\mathrm{Total}$ & $\es{1.45}{-4}$  & $\es{9.69}{-5}$  & $\es{2.26}{-2}$  & $\es{6.18}{-2}$  & $\es{2.02}{-1}$  & $\es{1.83}{-4}$   & $\es{8.05}{-4}$  & $\es{2.14}{-2}$  \\
\hline
                 & M15bE0.82        & M15bE0.89        & M15bE0.92        & M15bE1.69        & M15bE2.63        & M15bE10.7        & M15cE2.06         & M15cE1.94        \\  
\hline
$\mathrm{C}$     & $\es{3.87}{-3}$  & $\es{4.34}{-2}$  & $\es{4.31}{-2}$  & $\es{4.47}{-2}$  & $\es{4.48}{-2}$  & $\es{4.47}{-2}$  & $\es{4.47}{-2}$   & $\es{4.47}{-2}$  \\
$\mathrm{SiC}$   & $\es{3.37}{-12}$ & $\es{5.84}{-12}$ & $\es{6.10}{-12}$ & $\es{2.11}{-12}$ & $\es{3.56}{-10}$ & $\es{6.54}{-12}$ & $\es{1.13}{-12}$  & $\es{1.24}{-12}$ \\
$\mathrm{TiC}$   & $\es{1.28}{-12}$ & $\es{3.97}{-12}$ & $\es{4.89}{-12}$ & $\es{2.61}{-12}$ & $\es{1.96}{-12}$ & $\es{9.59}{-12}$ & $\es{1.17}{-12}$  & $\es{1.29}{-12}$ \\
$\mathrm{Si}$    &  0               &  0               &  0               & $\es{2.87}{-11}$ & $\es{7.98}{-8}$  & $\es{3.66}{-28}$ & $\es{6.31}{-10}$  & $\es{1.20}{-10}$ \\
$\silica$        & $\es{3.12}{-7}$  & $\es{2.07}{-7}$  & $\es{2.00}{-7}$  & $\es{9.71}{-8}$  & $\es{5.74}{-8}$  & $\es{1.72}{-7}$  & $\es{7.36}{-8}$   & $\es{7.90}{-8}$  \\
$\enstatite$     & $\es{1.75}{-6}$  & $\es{1.16}{-6}$  & $\es{1.12}{-6}$  & $\es{5.57}{-7}$  & $\es{5.54}{-4}$  & $\es{9.74}{-7}$  & $\es{1.83}{-6}$   & $\es{1.47}{-6}$  \\
$\forsterite$    & $\es{1.37}{-5}$  & $\es{9.17}{-6}$  & $\es{8.84}{-6}$  & $\es{4.82}{-6}$  & $\es{1.20}{-1}$  & $\es{7.65}{-6}$  & $\es{7.90}{-2}$   & $\es{6.43}{-2}$  \\
$\magnesia$      & $\es{3.43}{-8}$  & $\es{2.29}{-8}$  & $\es{2.21}{-8}$  & $\es{1.05}{-8}$  & $\es{6.11}{-9}$  & $\es{1.89}{-8}$  & $\es{7.90}{-9}$   & $\es{8.49}{-9}$  \\
$\alumina$       & $\es{6.42}{-7}$  & $\es{4.36}{-7}$  & $\es{4.21}{-7}$  & $\es{1.89}{-2}$  & $\es{1.74}{-2}$  & $\es{1.57}{-4}$  & $\es{1.83}{-2}$   & $\es{1.84}{-2}$  \\
$\mathrm{FeO}$   & $\es{4.30}{-6}$  & $\es{2.99}{-6}$  & $\es{2.89}{-6}$  & $\es{1.46}{-6}$  & $\es{8.85}{-7}$  & $\es{2.51}{-6}$  & $\es{1.12}{-6}$   & $\es{1.20}{-6}$  \\
$\magnetite$     & $\es{6.02}{-5}$  & $\es{4.14}{-5}$  & $\es{3.99}{-5}$  & $\es{2.04}{-5}$  & $\es{1.25}{-5}$  & $\es{3.48}{-5}$  & $\es{1.58}{-5}$   & $\es{1.68}{-5}$  \\
$\mathrm{Total}$ & $\es{3.96}{-3}$  & $\es{4.34}{-2}$  & $\es{4.31}{-2}$  & $\es{6.37}{-2}$  & $\es{1.83}{-1}$  & $\es{4.49}{-2}$  & $\es{1.42}{-1}$   & $\es{1.27}{-1}$  \\
\hline
                 & M15cE1.90        & M15cE1.86        & M15cE2.24        & M15cE2.60        & M15cE3.43        & M20aE0.53        & M20aE0.65         & M20aE0.81        \\  
\hline
$\mathrm{C}$     & $\es{4.47}{-2}$  & $\es{4.47}{-2}$  & $\es{4.47}{-2}$  & $\es{4.47}{-2}$  & $\es{4.48}{-2}$  & $\es{2.36}{-2}$  & $\es{2.37}{-2}$   & $\es{2.37}{-2}$  \\ 
$\mathrm{SiC}$   & $\es{1.26}{-12}$ & $\es{1.34}{-12}$ & $\es{1.21}{-12}$ & $\es{4.86}{-10}$ & $\es{3.61}{-10}$ & $\es{3.72}{-12}$ & $\es{5.40}{-12}$  & $\es{6.39}{-12}$ \\ 
$\mathrm{TiC}$   & $\es{1.32}{-12}$ & $\es{1.41}{-12}$ & $\es{1.06}{-12}$ & $\es{8.59}{-13}$ & $\es{6.09}{-13}$ & $\es{4.57}{-17}$ & $\es{6.76}{-13}$  & $\es{1.03}{-11}$ \\ 
$\mathrm{Si}$    & $\es{5.04}{-11}$ & $\es{1.40}{-11}$ & $\es{2.67}{-9}$  & $\es{1.40}{-8}$  & $\es{1.29}{-8}$  &  0               &  0                &  0               \\ 
$\silica$        & $\es{8.09}{-8}$  & $\es{8.60}{-8}$  & $\es{6.66}{-8}$  & $\es{5.70}{-8}$  & $\es{1.11}{-2}$  & $\es{4.53}{-7}$  & $\es{3.84}{-7}$   & $\es{3.34}{-7}$  \\
$\enstatite$     & $\es{1.08}{-6}$  & $\es{5.49}{-7}$  & $\es{1.46}{-4}$  & $\es{4.15}{-4}$  & $\es{7.02}{-4}$  & $\es{2.68}{-6}$  & $\es{2.35}{-6}$   & $\es{1.85}{-6}$  \\
$\forsterite$    & $\es{4.06}{-2}$  & $\es{1.41}{-3}$  & $\es{1.15}{-1}$  & $\es{1.28}{-1}$  & $\es{1.28}{-1}$  & $\es{2.11}{-5}$  & $\es{1.88}{-5}$   & $\es{1.47}{-5}$  \\
$\magnesia$      & $\es{8.71}{-9}$  & $\es{9.27}{-9}$  & $\es{7.13}{-9}$  & $\es{6.03}{-9}$  & $\es{2.75}{-7}$  & $\es{4.88}{-8}$  & $\es{4.16}{-8}$   & $\es{3.42}{-8}$  \\
$\alumina$       & $\es{1.84}{-2}$  & $\es{1.84}{-2}$  & $\es{1.83}{-2}$  & $\es{1.82}{-2}$  & $\es{1.82}{-2}$  & $\es{1.00}{-6}$  & $\es{8.24}{-7}$   & $\es{1.83}{-3}$  \\ 
$\mathrm{FeO}$   & $\es{1.22}{-6}$  & $\es{1.30}{-6}$  & $\es{1.02}{-6}$  & $\es{8.61}{-7}$  & $\es{6.86}{-7}$  & $\es{5.91}{-6}$  & $\es{5.17}{-6}$   & $\es{4.36}{-6}$  \\ 
$\magnetite$     & $\es{1.72}{-5}$  & $\es{1.82}{-5}$  & $\es{1.43}{-5}$  & $\es{1.24}{-5}$  & $\es{1.04}{-5}$  & $\es{8.39}{-5}$  & $\es{7.28}{-5}$   & $\es{6.31}{-5}$  \\
$\mathrm{Total}$ & $\es{1.03}{-1}$  & $\es{6.46}{-2}$  & $\es{1.78}{-1}$  & $\es{1.92}{-1}$  & $\es{2.03}{-1}$  & $\es{2.37}{-2}$  & $\es{2.38}{-2}$   & $\es{2.56}{-2}$  \\
\hline
                 & M20aE0.85         & M20aE1.39        & M20aE1.47        & M20aE2.43         & M20aE2.50         & M20aE4.15         & M20bE0.78          & M20bE1.04   \\  
\hline
$\mathrm{C}$     & $\es{2.37}{-2}$  & $\es{2.37}{-2}$  & $\es{2.37}{-2}$  & $\es{2.37}{-2}$  & $\es{2.37}{-2}$  & $\es{2.38}{-2}$  & $\es{2.37}{-2}$   & $\es{2.37}{-2}$  \\ 
$\mathrm{SiC}$   & $\es{6.53}{-12}$ & $\es{6.17}{-12}$ & $\es{4.88}{-12}$ & $\es{3.55}{-12}$ & $\es{3.57}{-12}$ & $\es{2.60}{-12}$ & $\es{6.52}{-12}$  & $\es{6.19}{-12}$ \\ 
$\mathrm{TiC}$   & $\es{1.10}{-11}$ & $\es{1.41}{-11}$ & $\es{9.82}{-12}$ & $\es{8.09}{-12}$ & $\es{7.95}{-12}$ & $\es{5.89}{-12}$ & $\es{1.08}{-11}$  & $\es{1.36}{-11}$ \\ 
$\mathrm{Si}$    &  0               & $\es{1.34}{-15}$ & $\es{1.86}{-11}$ & $\es{5.35}{-7}$  & $\es{3.15}{-7}$  & $\es{2.85}{-7}$  &  0                & $\es{1.12}{-30}$ \\ 
$\silica$        & $\es{3.18}{-7}$  & $\es{3.23}{-7}$  & $\es{3.20}{-7}$  & $\es{2.51}{-7}$  & $\es{2.47}{-7}$  & $\es{2.90}{-2}$  & $\es{3.34}{-7}$   & $\es{2.52}{-7}$  \\
$\enstatite$     & $\es{1.75}{-6}$  & $\es{4.36}{-6}$  & $\es{4.14}{-6}$  & $\es{3.40}{-6}$  & $\es{3.48}{-6}$  & $\es{8.33}{-4}$  & $\es{1.94}{-6}$   & $\es{5.05}{-6}$  \\
$\forsterite$    & $\es{1.39}{-5}$  & $\es{4.06}{-2}$  & $\es{3.00}{-2}$  & $\es{6.10}{-2}$  & $\es{5.62}{-2}$  & $\es{8.59}{-2}$  & $\es{1.54}{-5}$   & $\es{4.28}{-3}$  \\
$\magnesia$      & $\es{3.31}{-8}$  & $\es{1.96}{-7}$  & $\es{5.85}{-7}$  & $\es{6.71}{-7}$  & $\es{6.93}{-7}$  & $\es{8.43}{-7}$  & $\es{3.55}{-8}$   & $\es{2.78}{-8}$  \\
$\alumina$       & $\es{3.88}{-3}$  & $\es{1.85}{-2}$  & $\es{1.34}{-2}$  & $\es{1.15}{-2}$  & $\es{8.38}{-3}$  & $\es{9.92}{-3}$  & $\es{1.27}{-3}$   & $\es{7.87}{-3}$  \\ 
$\mathrm{FeO}$   & $\es{4.24}{-6}$  & $\es{2.63}{-6}$  & $\es{4.40}{-6}$  & $\es{3.31}{-6}$  & $\es{3.27}{-6}$  & $\es{2.37}{-6}$  & $\es{4.51}{-6}$   & $\es{3.60}{-6}$  \\ 
$\magnetite$     & $\es{6.02}{-5}$  & $\es{6.80}{-5}$  & $\es{6.75}{-5}$  & $\es{5.26}{-5}$  & $\es{5.24}{-5}$  & $\es{3.89}{-5}$  & $\es{6.49}{-5}$   & $\es{4.88}{-5}$  \\
$\mathrm{Total}$ & $\es{2.77}{-2}$  & $\es{8.30}{-2}$  & $\es{6.73}{-2}$  & $\es{9.64}{-2}$  & $\es{8.84}{-2}$  & $\es{1.49}{-1}$  & $\es{2.51}{-2}$   & $\es{3.59}{-2}$  \\
\hline
                 & M20bE1.19        & M20bE1.52        & M20bE2.60        & M20bE4.33        & M20cE0.75        & M20cE0.84        & M20cE1.00         & M20cE1.65        \\  
\hline
$\mathrm{C}$     & $\es{2.37}{-2}$  & $\es{2.37}{-2}$  & $\es{2.37}{-2}$  & $\es{2.38}{-2}$  & $\es{2.37}{-2}$  & $\es{2.37}{-2}$  & $\es{2.37}{-2}$   & $\es{2.37}{-2}$  \\ 
$\mathrm{SiC}$   & $\es{5.75}{-12}$ & $\es{5.55}{-12}$ & $\es{3.28}{-12}$ & $\es{2.46}{-12}$ & $\es{6.45}{-12}$ & $\es{6.73}{-12}$ & $\es{5.41}{-12}$  & $\es{5.72}{-12}$ \\ 
$\mathrm{TiC}$   & $\es{1.24}{-11}$ & $\es{1.27}{-11}$ & $\es{7.34}{-12}$ & $\es{5.61}{-12}$ & $\es{9.94}{-12}$ & $\es{1.13}{-11}$ & $\es{1.07}{-11}$  & $\es{1.45}{-11}$ \\ 
$\mathrm{Si}$    & $\es{2.23}{-19}$ & $\es{2.46}{-13}$ & $\es{4.49}{-7}$  & $\es{2.64}{-7}$  &  0               &  0               & $\es{7.82}{-35}$  & $\es{3.11}{-9}$  \\ 
$\silica$        & $\es{2.15}{-7}$  & $\es{3.14}{-7}$  & $\es{2.41}{-7}$  & $\es{1.81}{-2}$  & $\es{3.37}{-7}$  & $\es{3.24}{-7}$  & $\es{2.70}{-7}$   & $\es{3.02}{-7}$  \\
$\enstatite$     & $\es{4.63}{-6}$  & $\es{4.11}{-6}$  & $\es{3.37}{-6}$  & $\es{1.02}{-3}$  & $\es{1.97}{-6}$  & $\es{1.78}{-6}$  & $\es{5.22}{-6}$   & $\es{3.92}{-6}$  \\
$\forsterite$    & $\es{1.47}{-2}$  & $\es{4.30}{-2}$  & $\es{6.13}{-2}$  & $\es{8.71}{-2}$  & $\es{1.57}{-5}$  & $\es{1.41}{-5}$  & $\es{4.44}{-3}$   & $\es{5.35}{-2}$  \\
$\magnesia$      & $\es{2.36}{-8}$  & $\es{6.02}{-7}$  & $\es{6.99}{-7}$  & $\es{8.22}{-7}$  & $\es{3.60}{-8}$  & $\es{3.33}{-8}$  & $\es{2.98}{-8}$   & $\es{6.08}{-7}$  \\
$\alumina$       & $\es{1.14}{-2}$  & $\es{1.80}{-2}$  & $\es{1.09}{-2}$  & $\es{8.66}{-3}$  & $\es{5.62}{-4}$  & $\es{3.46}{-3}$  & $\es{9.61}{-3}$   & $\es{1.92}{-2}$  \\ 
$\mathrm{FeO}$   & $\es{3.11}{-6}$  & $\es{4.33}{-6}$  & $\es{3.18}{-6}$  & $\es{2.27}{-6}$  & $\es{4.56}{-6}$  & $\es{4.26}{-6}$  & $\es{3.81}{-6}$   & $\es{4.14}{-6}$  \\ 
$\magnetite$     & $\es{4.25}{-5}$  & $\es{6.67}{-5}$  & $\es{5.08}{-5}$  & $\es{3.72}{-5}$  & $\es{6.56}{-5}$  & $\es{6.12}{-5}$  & $\es{5.16}{-5}$   & $\es{6.40}{-5}$  \\
$\mathrm{Total}$ & $\es{4.99}{-2}$  & $\es{8.49}{-2}$  & $\es{9.62}{-2}$  & $\es{1.38}{-1}$  & $\es{2.43}{-2}$  & $\es{2.72}{-2}$  & $\es{3.78}{-2}$   & $\es{9.66}{-2}$  \\
\hline
                 & M20cE2.76        & M20cE2.85        & M20cE5.03        & M20cE8.86        & M20dE4.3        & M20dE5.9        & M20dE18.1        &  M20dE64.5          \\  
\hline
$\mathrm{C}$     & $\es{2.37}{-2}$  & $\es{2.37}{-2}$  & $\es{2.38}{-2}$  & $\es{2.42}{-2}$  & $\es{1.47}{-2}$  & $\es{2.30}{-2}$  & $\es{2.37}{-2}$   & $\es{2.36}{-2}$  \\ 
$\mathrm{SiC}$   & $\es{3.30}{-12}$ & $\es{3.30}{-12}$ & $\es{9.78}{-11}$ & $\es{3.02}{-10}$ & $\es{5.73}{-13}$ & $\es{2.50}{-13}$ & $\es{3.16}{-13}$  & $\es{7.28}{-14}$ \\ 
$\mathrm{TiC}$   & $\es{7.45}{-12}$ & $\es{6.80}{-12}$ & $\es{4.97}{-12}$ & $\es{3.49}{-12}$ & $\es{1.59}{-12}$ & $\es{6.30}{-13}$ & $\es{5.01}{-13}$  & $\es{1.47}{-13}$ \\ 
$\mathrm{Si}$    & $\es{4.72}{-7}$  & $\es{4.29}{-7}$  & $\es{2.02}{-7}$  & $\es{5.47}{-4}$  & $\es{3.11}{-12}$ & $\es{4.39}{-11}$ & $\es{5.86}{-9}$   & $\es{3.58}{-9}$  \\ 
$\silica$        & $\es{2.33}{-7}$  & $\es{7.95}{-7}$  & $\es{2.50}{-2}$  & $\es{3.21}{-3}$  & $\es{1.02}{-1}$  & $\es{9.35}{-8}$  & $\es{1.64}{-7}$   & $\es{1.45}{-7}$  \\
$\enstatite$     & $\es{2.41}{-5}$  & $\es{2.32}{-5}$  & $\es{1.02}{-3}$  & $\es{6.83}{-3}$  & $\es{2.41}{-5}$  & $\es{3.64}{-6}$  & $\es{2.53}{-6}$   & $\es{2.22}{-6}$  \\
$\forsterite$    & $\es{6.89}{-2}$  & $\es{6.86}{-2}$  & $\es{8.70}{-2}$  & $\es{8.76}{-2}$  & $\es{6.44}{-2}$  & $\es{4.64}{-2}$  & $\es{2.11}{-2}$   & $\es{1.22}{-2}$  \\
$\magnesia$      & $\es{6.59}{-7}$  & $\es{7.35}{-7}$  & $\es{7.60}{-7}$  & $\es{2.02}{-6}$  & $\es{1.73}{-6}$  & $\es{5.26}{-6}$  & $\es{6.62}{-7}$   & $\es{5.76}{-7}$  \\
$\alumina$       & $\es{1.19}{-2}$  & $\es{1.09}{-2}$  & $\es{8.58}{-3}$  & $\es{3.63}{-3}$  & $\es{2.64}{-3}$  & $\es{7.47}{-3}$  & $\es{5.58}{-3}$   & $\es{2.46}{-3}$  \\ 
$\mathrm{FeO}$   & $\es{3.05}{-6}$  & $\es{3.05}{-6}$  & $\es{2.01}{-6}$  & $\es{1.36}{-6}$  & $\es{7.79}{-7}$  & $\es{1.49}{-6}$  & $\es{2.08}{-6}$   & $\es{1.72}{-6}$  \\ 
$\magnetite$     & $\es{4.88}{-5}$  & $\es{4.90}{-5}$  & $\es{3.31}{-5}$  & $\es{2.32}{-5}$  & $\es{1.42}{-5}$  & $\es{2.78}{-5}$  & $\es{3.57}{-5}$   & $\es{2.95}{-5}$  \\
$\mathrm{Total}$ & $\es{1.04}{-1}$  & $\es{1.03}{-1}$  & $\es{1.45}{-1}$  & $\es{1.31}{-1}$  & $\es{1.84}{-1}$  & $\es{7.70}{-2}$  & $\es{5.05}{-2}$   & $\es{3.84}{-2}$  \\
\hline
                 & M20E78.9          & M20dE88.4        & M20dE124.0       & M25aE0.99        & M25aE1.57        & M25aE4.73        & M25aE6.17         & M25aE7.42        \\  
\hline
$\mathrm{C}$     &  $\es{2.00}{-2}$               & $\es{1.44}{-2}$  & $\es{1.12}{-2}$  & $\es{1.37}{-2}$  & $\es{1.37}{-2}$  & $\es{1.38}{-2}$  & $\es{1.39}{-2}$   & $\es{1.39}{-2}$  \\ 
$\mathrm{SiC}$   &  $\es{1.48}{-12}$               & $\es{3.84}{-13}$ & $\es{4.61}{-14}$ & $\es{7.82}{-13}$ & $\es{4.63}{-13}$ & $\es{2.09}{-13}$ & $\es{1.56}{-13}$  & $\es{1.37}{-13}$ \\ 
$\mathrm{TiC}$   &  $\es{4.62}{-13}$               & $\es{1.08}{-12}$ & $\es{1.00}{-13}$ & $\es{1.96}{-12}$ & $\es{1.15}{-12}$ & $\es{7.05}{-13}$ & $\es{4.86}{-13}$  & $\es{4.67}{-13}$ \\ 
$\mathrm{Si}$    &  $\es{8.61}{-9}$               & $\es{2.68}{-12}$ & $\es{2.26}{-15}$ & $\es{4.86}{-6}$  & $\es{7.67}{-7}$  & $\es{1.34}{-1}$  & $\es{1.16}{-1}$   & $\es{1.22}{-1}$  \\ 
$\silica$        &  $\es{1.51}{-4}$               & $\es{1.24}{-4}$  & $\es{1.20}{-2}$  & $\es{5.92}{-8}$  & $\es{4.26}{-8}$  & $\es{4.19}{-2}$  & $\es{5.43}{-2}$   & $\es{4.59}{-2}$  \\
$\enstatite$     &  $\es{3.18}{-6}$               & $\es{8.09}{-5}$  & $\es{2.18}{-5}$  & $\es{7.67}{-7}$  & $\es{1.11}{-6}$  & $\es{1.13}{-3}$  & $\es{2.26}{-3}$   & $\es{3.05}{-3}$  \\
$\forsterite$    &  $\es{1.54}{-2}$               & $\es{4.50}{-2}$  & $\es{1.69}{-3}$  & $\es{3.59}{-2}$  & $\es{6.15}{-2}$  & $\es{1.60}{-1}$  & $\es{2.62}{-1}$   & $\es{2.77}{-1}$  \\
$\magnesia$      &  $\es{4.21}{-2}$               & $\es{2.48}{-3}$  & $\es{7.47}{-6}$  & $\es{1.20}{-7}$  & $\es{8.15}{-8}$  & $\es{8.00}{-8}$  & $\es{5.18}{-7}$   & $\es{7.36}{-6}$  \\
$\alumina$       &  $\es{4.93}{-3}$               & $\es{4.37}{-3}$  & $\es{7.11}{-5}$  & $\es{9.32}{-3}$  & $\es{1.42}{-2}$  & $\es{1.91}{-2}$  & $\es{2.41}{-2}$   & $\es{1.97}{-2}$  \\ 
$\mathrm{FeO}$   &  $\es{3.76}{-5}$               & $\es{1.06}{-6}$  & $\es{3.72}{-7}$  & $\es{7.86}{-7}$  & $\es{5.55}{-7}$  & $\es{1.01}{-6}$  & $\es{8.11}{-7}$   & $\es{9.69}{-7}$  \\ 
$\magnetite$     &  $\es{1.51}{-3}$               & $\es{1.55}{-5}$  & $\es{7.91}{-6}$  & $\es{1.16}{-5}$  & $\es{8.30}{-6}$  & $\es{1.49}{-5}$  & $\es{1.31}{-5}$   & $\es{1.80}{-5}$  \\
$\mathrm{Total}$ &  $\es{8.45}{-2}$               & $\es{6.65}{-2}$  & $\es{2.51}{-2}$  & $\es{5.90}{-2}$  & $\es{8.95}{-2}$  & $\es{4.76}{-1}$  & $\es{5.21}{-1}$   & $\es{5.31}{-1}$  \\
\hline
                 & M25aE14.8        & M25bE8.40        & M25bE9.73        & M25bE18.4        & M25d3E0.89       & M25d3E0.92       & M25d3E1.04        & M25d3E1.20       \\  
\hline
$\mathrm{C}$     & $\es{1.48}{-2}$  & $\es{1.39}{-2}$  & $\es{1.39}{-2}$  & $\es{1.40}{-2}$  & $\es{1.37}{-2}$  & $\es{1.37}{-2}$  & $\es{3.60}{-3}$   & $\es{1.98}{-3}$  \\ 
$\mathrm{SiC}$   & $\es{1.13}{-13}$ & $\es{1.71}{-13}$ & $\es{1.93}{-13}$ & $\es{5.79}{-14}$ & $\es{4.97}{-13}$ & $\es{9.73}{-11}$ & $\es{8.99}{-10}$  & $\es{1.63}{-9}$  \\ 
$\mathrm{TiC}$   & $\es{4.99}{-13}$ & $\es{6.13}{-13}$ & $\es{6.77}{-13}$ & $\es{1.65}{-13}$ & $\es{1.30}{-12}$ & $\es{5.97}{-13}$ & $\es{3.99}{-12}$  & $\es{1.36}{-12}$ \\ 
$\mathrm{Si}$    & $\es{1.78}{-1}$  & $\es{3.43}{-5}$  & $\es{8.73}{-4}$  & $\es{5.31}{-5}$  & $\es{3.58}{-6}$  & $\es{1.20}{-1}$  & $\es{1.20}{-1}$   & $\es{1.17}{-1}$  \\ 
$\silica$        & $\es{9.06}{-2}$  & $\es{8.47}{-2}$  & $\es{1.11}{-1}$  & $\es{2.12}{-2}$  & $\es{5.93}{-8}$  & $\es{2.34}{-1}$  & $\es{5.57}{-2}$   & $\es{2.82}{-2}$  \\
$\enstatite$     & $\es{7.19}{-3}$  & $\es{3.91}{-2}$  & $\es{1.10}{-3}$  & $\es{2.60}{-3}$  & $\es{7.11}{-7}$  & $\es{1.50}{-3}$  & $\es{2.27}{-2}$   & $\es{1.23}{-3}$  \\
$\forsterite$    & $\es{8.90}{-2}$  & $\es{2.46}{-1}$  & $\es{2.56}{-1}$  & $\es{1.03}{-1}$  & $\es{3.68}{-2}$  & $\es{2.08}{-1}$  & $\es{1.73}{-1}$   & $\es{2.57}{-2}$  \\
$\magnesia$      & $\es{1.35}{-4}$  & $\es{8.59}{-6}$  & $\es{3.89}{-10}$ & $\es{7.67}{-5}$  & $\es{1.07}{-7}$  & $\es{5.08}{-8}$  & $\es{3.27}{-5}$   & $\es{8.42}{-5}$  \\ 
$\alumina$       & $\es{2.30}{-3}$  & $\es{6.48}{-3}$  & $\es{5.29}{-2}$  & $\es{1.62}{-3}$  & $\es{9.70}{-3}$  & $\es{4.82}{-2}$  & $\es{2.51}{-3}$   & $\es{9.10}{-4}$  \\ 
$\mathrm{FeO}$   & $\es{1.43}{-6}$  & $\es{3.11}{-7}$  & $\es{1.70}{-7}$  & $\es{1.12}{-6}$  & $\es{7.91}{-7}$  & $\es{2.17}{-6}$  & $\es{8.58}{-7}$   & $\es{1.24}{-6}$  \\ 
$\magnetite$     & $\es{3.05}{-5}$  & $\es{5.68}{-6}$  & $\es{3.81}{-6}$  & $\es{2.22}{-5}$  & $\es{1.16}{-5}$  & $\es{4.03}{-5}$  & $\es{1.76}{-5}$   & $\es{2.65}{-5}$  \\ 
$\mathrm{Total}$ & $\es{3.88}{-1}$  & $\es{3.91}{-1}$  & $\es{4.36}{-1}$  & $\es{1.43}{-1}$  & $\es{6.03}{-2}$  & $\es{6.93}{-1}$  & $\es{5.38}{-1}$   & $\es{5.31}{-1}$  \\ 
\hline
                 & M25d2E2.53       & M25d2E2.64       & M25d2E2.78       & M25d2E3.07       & M25d1E3.30       & M25d1E4.72       & M25d1E7.08       \\ 
\hline
$\mathrm{C}$     & $\es{1.37}{-2}$  & $\es{1.37}{-2}$  & $\es{1.37}{-2}$  & $\es{4.44}{-3}$  & $\es{1.37}{-2}$  & $\es{1.37}{-2}$  & $\es{1.37}{-2}$  \\
$\mathrm{SiC}$   & $\es{1.54}{-12}$ & $\es{1.28}{-12}$ & $\es{8.54}{-13}$ & $\es{7.19}{-11}$ & $\es{6.65}{-13}$ & $\es{4.26}{-13}$ & $\es{1.74}{-11}$ \\
$\mathrm{TiC}$   & $\es{5.57}{-12}$ & $\es{4.11}{-12}$ & $\es{2.12}{-12}$ & $\es{3.34}{-12}$ & $\es{1.70}{-12}$ & $\es{1.29}{-12}$ & $\es{6.07}{-13}$ \\
$\mathrm{Si}$    & $\es{8.12}{-3}$  & $\es{7.20}{-2}$  & $\es{7.20}{-2}$  & $\es{1.39}{-1}$  & $\es{7.88}{-2}$  & $\es{7.29}{-2}$  & $\es{7.26}{-2}$  \\
$\silica$        & $\es{3.05}{-1}$  & $\es{3.29}{-1}$  & $\es{1.85}{-1}$  & $\es{5.76}{-2}$  & $\es{1.26}{-1}$  & $\es{3.40}{-1}$  & $\es{3.38}{-1}$  \\
$\enstatite$     & $\es{5.58}{-4}$  & $\es{1.07}{-4}$  & $\es{2.62}{-3}$  & $\es{1.07}{-3}$  & $\es{1.96}{-3}$  & $\es{5.35}{-5}$  & $\es{6.67}{-5}$  \\
$\forsterite$    & $\es{2.53}{-1}$  & $\es{2.52}{-1}$  & $\es{2.52}{-1}$  & $\es{1.98}{-2}$  & $\es{2.52}{-1}$  & $\es{2.52}{-1}$  & $\es{2.52}{-1}$  \\
$\magnesia$      & $\es{2.13}{-7}$  & $\es{5.97}{-8}$  & $\es{1.43}{-9}$  & $\es{6.74}{-5}$  & $\es{1.45}{-9}$  & $\es{9.97}{-10}$ & $\es{7.16}{-3}$  \\
$\alumina$       & $\es{4.28}{-2}$  & $\es{4.42}{-2}$  & $\es{4.78}{-2}$  & $\es{7.86}{-4}$  & $\es{5.58}{-2}$  & $\es{5.70}{-2}$  & $\es{5.70}{-2}$  \\
$\mathrm{FeO}$   & $\es{1.20}{-5}$  & $\es{8.67}{-6}$  & $\es{2.11}{-6}$  & $\es{1.38}{-6}$  & $\es{7.25}{-7}$  & $\es{9.78}{-6}$  & $\es{1.02}{-5}$  \\
$\magnetite$     & $\es{5.68}{-4}$  & $\es{7.39}{-4}$  & $\es{3.91}{-5}$  & $\es{2.90}{-5}$  & $\es{1.23}{-5}$  & $\es{1.07}{-3}$  & $\es{1.72}{-3}$  \\
$\mathrm{Total}$ & $\es{6.26}{-1}$  & $\es{7.35}{-1}$  & $\es{5.89}{-1}$  & $\es{3.05}{-1}$  & $\es{5.31}{-1}$  & $\es{7.40}{-1}$  & $\es{7.68}{-1}$  \\
\enddata
\tablecomments{The single element Fe-group grain species are excluded from this table due to very inconsistent non-negligible yields. Entries with a dash line indicate negligible and effectively zero amounts of that species being produced for the given model.}
\end{deluxetable}